\documentclass[journel=jctcce, manuscript=article, layout=twocolumn]{achemso}
\setkeys{acs}{maxauthors=0, etalmode=truncate}

\usepackage{graphicx}
\usepackage{subcaption}
\usepackage{amsmath}
\usepackage{bm}
\usepackage{braket}
\newcommand\mydots{\makebox[1em][c]{.\hfil.\hfil.}}
\usepackage{multirow}
\usepackage{booktabs}
\usepackage{float}
\usepackage[title]{appendix}
\DeclareMathAlphabet{\mathpzc}{OT1}{pzc}{m}{it}
\usepackage{listings}
\usepackage[linenumbers=true, charsperline=44, theme=default-plain]{jlcode}
\mciteErrorOnUnknownfalse
\SectionNumbersOn

\author{Weishi Wang}
\email{weishi.wang.gr@dartmouth.edu}
\affiliation{Department of Physics and Astronomy, Dartmouth College, Hanover, New Hampshire 03755, USA}

\author{James D. Whitfield}
\email{james.d.whitfield@dartmouth.edu}
\affiliation{Department of Physics and Astronomy, Dartmouth College, Hanover, New Hampshire 03755, USA}
\alsoaffiliation{AWS Center for Quantum Computing, Pasadena, California 91106, USA}

\title{Basis set generation and optimization in the NISQ era with Quiqbox.jl}

\begin{document}

\begin{abstract}
    In the noisy intermediate-scale quantum era, ab initio computation of the electronic structure problems has become one of the major benchmarks for identifying the boundary between classical and quantum computational power. Basis sets play a key role in the electronic structure methods implemented on both classical and quantum devices. To investigate the consequences of the single-particle basis sets, we propose a framework for more customizable basis set generation and optimization. This framework allows composite basis sets to go beyond typical basis set frameworks, such as atomic basis sets, by introducing the concept of mixed-contracted Gaussian-type orbitals. These basis set generations set the stage for more flexible variational optimization of basis set parameters. To realize this framework, we have developed an open-source software package named ``Quiqbox'' in the Julia programming language. We demonstrate various examples of using Quiqbox for basis set optimization and generation, ranging from optimizing atomic basis sets on the Hartree--Fock level, preparing the initial state for VQE computation, and constructing basis sets with completely delocalized orbitals. We also include various benchmarks of Quiqbox for basis set optimization and ab initial electronic structure computation.
\end{abstract}

% \maketitle

\section{Introduction}
Electronic structure~\cite{szabo2012modern, helgaker2013molecular} has been the intersection among multiple disciplines, including quantum physics, quantum chemistry, and material science. The goal of the field is to study the states and dynamics of many-electron systems, or more specifically, the electronic Hamiltonian~\cite{szabo2012modern} determined by a nuclear field. Consider a quantum system consisting of $N$ electrons and $N'$ nuclei, in which the $a'$th nucleus with charge $Z_{a'}$ is located on position ${\bm{R'_{a'}}}$. The information of all the nuclei forms a set that specifies the nuclear field of the system: 
\begin{equation}
    \mathbb{M}\equiv\{({\bm{R'_{a'}}}, Z_{a'})\,|\,a'\!=\!1,\dots ,N'\}.
\end{equation}
The electronic Hamiltonian of this system can be separated from its total Hamiltonian $H_{\rm tot}$ when the Born-Oppenheimer approximation~\cite{BornOppenheimerApp} is applied:
\begin{equation}
    H_{\rm elec}(\mathbb{M}; N^\uparrow, N^\downarrow) = H_{\rm tot}(\mathbb{M}; N) - H_{\rm nuc}(\mathbb{M})
\end{equation}
where $N^\uparrow$ and $N^\downarrow$ are the numbers of spin-up electrons and spin-down electrons, respectively, such that $N\!=\!N^\uparrow\!+\!N^\downarrow$. $H_{\rm nuc}$ is the sum of the nuclear kinetic energy and nuclear repulsion energy. Instead of being the direct sum of single-electron Hamiltonians, $H_{\rm elec}$ (in the Hartree atomic units) has the form: 
\begin{equation}
    \begin{aligned}
        &H_{\rm{elec}}(\mathbb{M}; N^\uparrow, N^\downarrow)\\
        =& \sum_{a=1}^N(-\frac{1}{2}\nabla_a^2\!-\!\!\sum_{a'=1}^{N'}\frac{Z_{a'}}{|\bm{r_a}\!-\!\bm{R'_{a'}}|}) + \hspace{-0.8em}\sum_{a=1, b>a}^{N,N}\!\frac{1}{|\bm{r_a}\!-\!\bm{r_b}|} \\
        % &\equiv \sum_{a}(-\frac{1}{2}\nabla_{\bm{A}}^2 - \sum_{A}\frac{Z_A}{r_{aA}}) + \sum_{b\neq a}\frac{1}{r_{ab}} \\
        \equiv& \sum_ah_1(a;\, \mathbb{M}) + \sum_{b>a}h_2(a,b)
        \label{eq:Helec}
    \end{aligned}    
\end{equation}
where $h_1(a;\, \mathbb{M})$ is a one-electron operator that represents the kinetic energy of the $a$th electron and the Coulomb attraction between it and the nuclei; $h_2(a, b)$ is a two-electron operator that represents the interaction between the $a$th and $b$th electrons. The sum of $h_1(a;\, \mathbb{M})$ forms the core Hamiltonian, and the sum of $h_2(a, b)$ makes up the electron--electron interaction in the system.

One of the basic goals of electronic structure study is to solve the eigenvalue problem of the electronic problem: 
\begin{equation}
    H_{\rm elec}\Phi_n = \mathcal{E}_n\Phi_n
\end{equation}
where $\Phi_n$ is the eigenfunction corresponding to the $n$th eigenenergy $\mathcal{E}_n$. $\Phi_n$ is a many-electron wavefunction that obeys the anti-symmetric constraint for identical fermions. 

In an effort to alleviate the exponentially growing computational cost with respect to the total number of electrons, various ansatzes and numerical optimizations for the many-electron eigenfunctions have been proposed. Within the domain of classical numerical simulation, mean-field methods such as the Hartree--Fock (HF) method~\cite{szabo2012modern, slater1951simplification, valatin1961generalized} were proposed to approximate the ground state with a non-interacting fermionic state (a single Slater determinant). Post-HF methods~\cite{szabo2012modern, bartlett1989alternative,  szalay2012multiconfiguration, mcardle2020quantum} try to further represent the ground state with the superposition of Slater determinants based on one reference state (HF state) or multiple reference states. 

With the successful demonstration on a quantum device to solve for the ground state of $\rm He$--$\rm{H^+}$~\cite{peruzzo_variational_2014}, variational quantum eigensolver (VQE)~\cite{mcclean2016theory, kandala2017hardware, cerezo_variational_2021, tilly2022variational} established its potential in utilizing noisy intermediate-scale quantum (NISQ) devices~\cite{shen2017quantum, preskill_quantum_2018, leymann2020bitter} for computing electronic structure. However, due to the limited scale of NISQ devices, the applicable basis sets to perform qubit encoding for VQE have been restricted to small-size basis sets~\cite{tilly2022variational}. On the other hand, the interest around basis set optimization for electronic structure has grown in the past few years~\cite{lehtola2015automatic, tamayo2018automatic, kasim2022dqc} with the development of computer algebra techniques such as automatic differentiation (AD)~\cite{griewank1989automatic, baydin2018automatic}.

We believe that a flexible control and comparable design of basis sets play a crucial role in developing and benchmarking electronic structure algorithms in the NISQ era when finding the boundary between classical and quantum computational power is vital for exploring practical quantum primacy~\cite{preskill_quantum_2018, daley2022practical, whitfield2022quantum}. Hence, we propose a unified framework for generating and optimizing various composite basis sets based on Gaussian-type orbital (GTO). This framework has been realized in an open-source software toolkit we developed, Quiqbox.jl (Quiqbox)~\cite{Quiqbox2022}, using the Julia programming language~\cite{Bezanson_Julia_A_fresh_2017, bezanson2018julia}. We show that with the help of Quiqbox, one can control, modify, and improve the existing basis sets composed of Gaussian-type atomic orbitals (AOs). Moreover, one can also construct and optimize more general basis functions beyond AOs. The basis set made of these basis functions can outperform atomic basis sets or achieve similar performance while meeting specific requirements such as minimal basis set size.

The paper is organized as follows. In Sec.~\ref{sec:intro-nisq}, we elaborate on the proposal to carry out a more systematic study of basis set design in the NISQ era. In Sec.~\ref{sec:bsDesign}, we describe the additional customizability our composite basis set generation framework brings for electronic structure computation and introduce a more general form of the linear combination of Gaussian-type orbitals (GTOs), mixed-contracted GTOs. We then derive the general expression of the basis set parameter gradient under the said framework and explain Quiqbox's optimization architecture in Sec.~\ref{sec:optPars}. From Sec.~\ref{sec:AbsOpt} to Sec.~\ref{sec:DO}, we demonstrate several examples of applying our framework for basis set generation and optimization using Quiqbox. We conclude the paper and discuss the outlook of potential basis set research directions in Sec.~\ref{sec:conclusion}. In Appx.~\ref{app:performance}, we present benchmarks on Quiqbox's functionalities that are mentioned in this paper and also offered by other electronic structure packages to provide an overview of Quiqbox's performance as a scientific software. Appx.~\ref{app:reaction} includes all the supplement information for Sec.~\ref{sec:H2O2}. Summaries of the essential acronyms and notation conventions we use throughout this paper are in the Appx.~\ref{app:notation}.

\section{Basis set design for the NISQ era}\label{sec:intro-nisq}

Let $E_0$ be the approximate ground-state energy of $H_{\rm elec}$ associated with a quantum many-electron system. Given an non-orthogonal basis set $\{\phi_n \,|\,  n=1,\dots,W\}$ that contains $W$ spatial orbitals with an overlap matrix $\bm{S}$ defined by $S_{ij}=\braket{\phi_i|\phi_j}$, we can construct an orthonormal basis set 
\begin{equation}
    \{\varphi_m\hspace{0.1em}|\hspace{0.1em}\forall m,n\hspace{-0.1em}\in\hspace{-0.1em}\{1,\dots,W\},\braket{\varphi_m|\varphi_n}\hspace{-0.15em}=\hspace{-0.15em}\delta_{mn}\}, 
\end{equation}
by performing a symmetric orthogonalization
\begin{equation}
    \varphi_m=\sum_nX_{nm}\phi_n, \;{\rm{with}}\; \bm{X} \equiv \bm{S}^{-1/2}.
    \label{eq:no2o}
\end{equation}
Thus, a trial state $\ket{\Psi_0(\bm{c})}$ parameterized by $\bm{c}$ can be constructed from $\{\varphi_m\}$ following an ansatz of the ground state. Depending on the selected ansatz, $\bm{c}$ may not only characterizes the single-electron modes but also contains the amplitudes of multiple Slater determinants formed by the modes. According to the variational principle, $E_0$ would be the lower bound of the energy expectation $\tilde{E_0}(\bm{c})$ with respect to $\ket{\Psi_0(\bm{c})}$:
\begin{equation}
    E_0 \leq \tilde{E_0}(\bm{c}) = \frac{\braket{\Psi_0(\bm{c})|H_{\rm elec}|\Psi_0(\bm{c})}}{\braket{\Psi_0(\bm{c})|\Psi_0(\bm{c})}}.
    \label{eq:E0v}
\end{equation}
Regardless of the choice of ansatz (e.g., the Hartree--Fock approximation) used to construct $\ket{\Psi_0(\bm{c})}$, $\tilde{E_0}(\bm{c})$ in Eq.~(\ref{eq:E0v}) can always be decomposed into three parts:
\begin{equation}
    \begin{aligned}
            \quad \bm{x} &= \bm{f_1}({\bm{A}}[\{\varphi_m\}]),\\
            \quad \bm{y} &= \bm{f_2}({\bm{B}}[\{\varphi_m\}]),\\
            \tilde{E_0}(\bm{c}) &= f_3(\bm{c};\, \bm{x},\, \bm{y}),
        \label{eq:E0general}
    \end{aligned}
\end{equation}
where vector functions $\bm{f_1}$: $\mathbb{R}^{W^2}\!\to\!\mathbb{R}^{W'}$ and $\bm{f_2}$: $\mathbb{R}^{W^4}\!\to\!\mathbb{R}^{W''}$ generate intermediate results $\bm{x}$ and $\bm{y}$ mapped from tensor ${\bm{A}}$ and tensor ${\bm{B}}$ which discretize the one-electron and two-electron operators respectively onto $\{\varphi_m\}$. The tensor elements of ${\bm{A}}$ and ${\bm{B}}$ are called electronic integrals (or molecular integrals in the context of molecular systems):
\begin{align}
    \begin{split}
        A_{mn} \hspace{-0.1em}&= \hspace{-0.5em}\int \hspace{-0.4em}d\bm{r_{\!a}}\varphi_m^*(\bm{r_{\!a}})h_1(a;\, \mathbb{M})\varphi_n(\bm{r_{\!a}})\\
    &\equiv (\varphi_m|\varphi_n), \label{eq:A}    
    \end{split}\\
    \begin{split}
    B_{mnpq} \hspace{-0.1em}&= \hspace{-0.5em}\int \hspace{-0.4em}d\bm{r_{\!a}}d\bm{r_b}\varphi_m^*\hspace{-0.1em}(\hspace{-0.1em}\bm{r_{\!a}}\hspace{-0.1em})\varphi_p^*\hspace{-0.1em}(\hspace{-0.1em}\bm{r_b}\hspace{-0.1em})h_2(\hspace{-0.1em}a,b\hspace{-0.1em})\varphi_q\hspace{-0.1em}(\hspace{-0.1em}\bm{r_b}\hspace{-0.1em})\varphi_n\hspace{-0.1em}(\hspace{-0.1em}\bm{r_{\!a}}\hspace{-0.1em})\\
    &\equiv (\varphi_m\varphi_n|\varphi_p\varphi_q).\label{eq:B}
    \end{split}
\end{align}
The third mapping function $f_3$ in Eq.~(\ref{eq:E0general}) takes $\bm{x}$ and $\bm{y}$ as parameters, and $\bm{c}$ as the input variable. In this way, the computation of $E_0$ is transformed into an optimization problem of finding the optimal value of $\bm{c}$:
\begin{equation}
    \bm{c^o} = \underset{\bm{c}}{{\rm arg}\,{\rm min}}\,\tilde{E_0}(\bm{c}).
    \label{eq:minimizeE}
\end{equation}
This can be done using either self-consistent field (SCF) methods~\cite{szabo2012modern, pulay1982improved, kudin2002black, hu2010accelerating}, or gradient-based methods realized by AD~\cite{yoshikawa2022automatic}.

One can see from Eq.~(\ref{eq:E0general}) that the expression of $\bm{f_1}$, $\bm{f_2}$, and $f_3$ depend on the choice of the ground state ansatz. More specifically, elements of $\bm{x}$ and $\bm{y}$ can be considered as contributions to the ground-state energy by the electronic integrals. An example of $\bm{f_1}$, $\bm{f_2}$ and $f_3$'s expression in the case of restricted closed-shell Hartree--Fock (RHF) method is shown in Eq.~(\ref{eq:dHF}) in Sec.~\ref{sec:generaldE}.

\begin{figure}[htbp]
    \centering
    \begin{subfigure}[b]{0.48\textwidth}
         \centering
         \includegraphics[width=\textwidth]{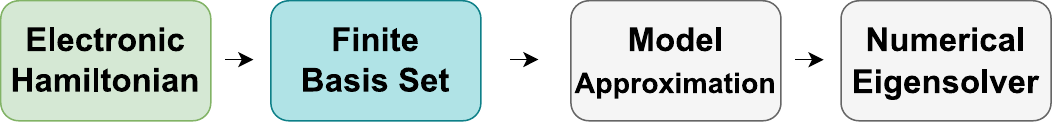}
         \caption{The general steps of solving a many-electron system.}
         \label{fig:ElectronicStructure1}
     \end{subfigure}
     \hfill
     \begin{subfigure}[b]{0.48\textwidth}
         \centering
         \includegraphics[width=\textwidth]{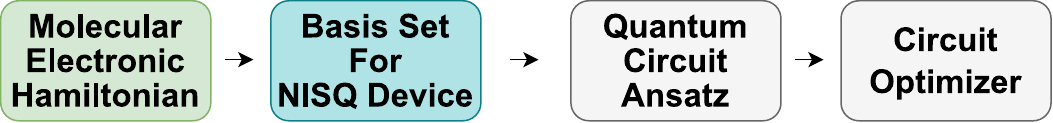}
         \caption{Applying VQE to solve the electronic structure of a molecular system.}
         \label{fig:ElectronicStructure2}
     \end{subfigure}
     \hfill
     \begin{subfigure}[b]{0.48\textwidth}
         \centering
         \includegraphics[width=\textwidth]{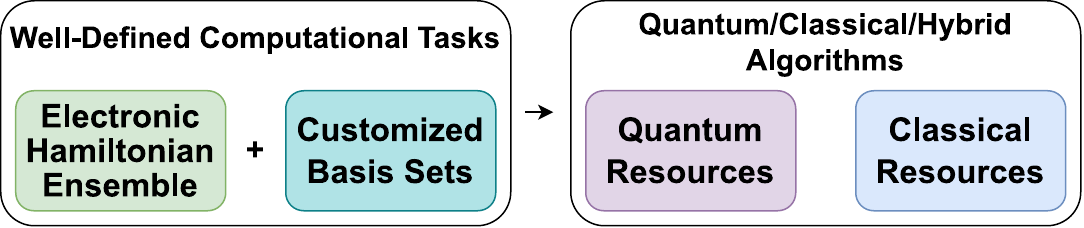}
         \caption{Using basis set design and molecular ensemble to form well-defined computational tasks for NISQ and classical algorithms.}
         \label{fig:ElectronicStructure3}
     \end{subfigure}
    \caption{The overall procedure of solving the electronic Hamiltonian can be divided into four steps for classical and quantum algorithms as shown in FIG.~\ref{fig:ElectronicStructure1} and FIG.~\ref{fig:ElectronicStructure2} respectively. A more systematic design and optimization of basis sets can aid the development of these algorithms for electronic structure problems, as shown in FIG.~\ref{fig:ElectronicStructure3}.}
    \label{fig:ElectronicStructure}
\end{figure}

Eqs.~(\ref{eq:A}) and (\ref{eq:B}) indicate that the discretization of $H_{\rm elec}$ is affected by choice of $\{\varphi_m\}$, which can introduce error due to the incompleteness of finite basis set and affect the hardness of the electronic structure problem~\cite{ogorman_electronic_2021}. Following Eqs.~(\ref{eq:E0general}--\ref{eq:minimizeE}), the process of solving for the electronic structure properties can be generalized into a standard procedure shown in FIG.~\ref{fig:ElectronicStructure1}. The electronic Hamiltonian of the system of interest is first discretized by a finite set of basis functions. After that, based on a model ansatz, a further approximation is applied (usually about the many-electron wavefunction) to construct the algebraic expression for the desired observable as an objective function (i.e., the ground-state energy) with respect to the discretized Hamiltonian. Thus, the objective function is implicitly a functional of the basis set. Finally, a numerical eigensolver (typically a variational optimizer) finds the approximate solution $\bm{c^o}$. Accordingly, the procedure of VQE for molecular electronic structure also falls into four similar steps as shown in FIG.~\ref{fig:ElectronicStructure2}.

From the generalized procedure of solving an electronic Hamiltonian, we showed why a basis set can affect both the accuracy and overhead of the overall computation. In the context of using VQE for electronic structure, this effect becomes more significant. Due to the limited number of qubits in NISQ devices for fermionic encoding~\cite{bravyi2002fermionic, nielsen2005fermionic, havlivcek2017operator, setia2019superfast}, there is often a trade-off between the size of the basis set and the size of the studied many-electron systems~\cite{mcardle2020quantum, aspuru2005simulated}. Moreover, the applied Gaussian-type AOs (e.g., STO-3G~\cite{stewart1970small}, cc-pVDZ~\cite{dunning1989gaussian}) are, in fact, not fully optimized for electronic structure properties such as the ground-state energy given a fixed number of single-electron modes. This can be supported by the literature about the optimization of AOs~\cite{tachikawa1999simultaneous, tamayo2018automatic}. For the purpose of improving the practicality of NISQ algorithms for electronic structure problems, we believe a systematic study of customizable basis set generation is of high importance. For instance, including basis set optimization and design as part of VQE's architecture can introduce the physical characteristics of many-electron systems in real space to VQE. This may improve various aspects of the algorithm, such as trial qubit state preparation or optimizer configuration for the landscape of the corresponding parameter space. To demonstrate the potential of a custom basis set for VQE, We include an example in Sec.~\ref{sec:VQE} of preparing an optimized basis set for ground-state VQE computation based on the framework we shall introduce in the next two sections.

Additionally, customizable basis set generation enables a better formulation to study the electronic Hamiltonian as a computational task. It has been shown that approximating the ground state of the electronic Hamiltonian with a fixed basis set in real space (and fixed particle number) is QMA-complete~\cite{ogorman_electronic_2021, bookatz2012qma}. This suggests that the way a basis set discretizes the real space can affect the hardness of the electronic structure problem. We can further incorporate this idea into basis set designs. Given an ensemble of electronic Hamiltonians with the same number of electrons, by designing customized basis sets, we shall generate instances of electronic structure problems classified by different computational complexities (see FIG.~\ref{fig:ElectronicStructure3}). Not only does this help study the complexity of electronic Hamiltonians, but it also brings in computational complexity as another tool for benchmarking quantum simulation algorithms~\cite{manin1980computable, feynman2018simulating, aspuru2005simulated, lanyon2010towards, cerezo_variational_2021, daley2022practical} against purely classical methods in the context of electronic structure.

\section{Higher customizability in Gaussian basis set design} \label{sec:bsDesign}
Basis sets used for electronic structure computation typically are classified into three categories: spatially delocalized basis sets that rely on periodic functions like plane waves~\cite{kresse1996efficient, kresse1996efficiency}, real-space basis sets that use localized functions such as wavelets distributed on grid points~\cite{pask1999real, genovese2008daubechies, ratcliff2020flexibilities}, and Gaussian basis sets that utilize GTOs to form contracted functions to approximate Slater-type orbitals~\cite{dunning1977gaussian, hill2013gaussian}. 

In our framework, we also use GTO as the building block of more complicated basis functions for several reasons. First and foremost, GTO currently is a common choice for constructing basis set for molecular systems, the scale of which is applicable to VQE in the current state of NISQ devices~\cite{peruzzo_variational_2014, mcardle2020quantum, tilly2022variational}. This allows us to optimize and build new basis sets on the existing progress of GTO research. Secondly, the orbital integrals from Eqs.~(\ref{eq:A}) and (\ref{eq:B}) in the case of GTO can be computed efficiently and analytically for the most part~\cite{head1988method, petersson2009detailed} so that the numerical error of the discretized electronic Hamiltonian can be controlled. Last but not least, by applying specific combinations of Gaussian functions, certain basis sets with different features from atomic basis sets, such as floating Gaussian basis sets~\cite{hurley1988computation, frost1967floating, huber1979geometry}, Gaussian-based real-space basis sets~\cite{white2017hybrid, white2019multisliced}, and even-tempered basis sets~\cite{cherkes2009spanning}, can also be constructed. 

In summary, by careful design, GTO can be used to generate more diverse basis functions beyond the existing Gaussian basis sets generation methods provided by the popular software libraries such as PySCF~\cite{sun2020recent} and Psi4~\cite{smith2020psi4} for specific needs. 

In this section, we first formally introduce GTO (Gaussian-type orbital). Then, we focus on describing the proposed framework that provides extra customizability in building GTO-based composite basis functions improving the versatility of Gaussian-based basis sets compared to the current standard approach. 

In general, the form of a GTO in Cartesian coordinates can be expressed as 
\begin{equation}
    \begin{alignedat}{2}
        & &&\phi^{\rm{GTO}}(\bm{r};\bm{R},\mathpzc{i},\mathpzc{j},\mathpzc{k},\alpha)\\
        &\,\equiv\,&& N(x\!-\!\!X)^{\mathpzc{i}}(y\!-\!\!Y)^{\mathpzc{j}}(z\!-\!\!Z)^{\mathpzc{k}}{\rm{exp}}(-\alpha|\bm{r\!-\!\!R}|^2),\\
        \mathpzc{l} &\equiv&& \mathpzc{i}\!+\!\mathpzc{j}\!+\!\mathpzc{k},\,
        % |\bm{r}| &=&& (x^2 + y^2 + z^2)^{\frac{1}{2}},\\
        % |\bm{R}| &=&& (X^2 + Y^2 + Z^2)^{\frac{1}{2}},
        \bm{r}\equiv[x,\, y,\, z],\,
        \bm{R}\equiv[X,\, Y,\, Z],
    \end{alignedat}
    \label{eq:GTO}
\end{equation}
where $\alpha$ is the exponent coefficient that determines how diffused the GTO is, $\mathpzc{l}$ represents the orbital angular momentum, and $\bm{R}$ is the center position of the orbital. The most common use of GTOs is to form a contracted Gaussian-type orbital (CGTO)
\begin{equation}
    \begin{aligned}
        &\phi_m^{\rm{CGTO}}(\bm{r}; \bm{R_m}, \mathpzc{i}_m, \mathpzc{j}_m, \mathpzc{k}_m)\\
        \equiv& \sum_n d_n\,\phi^{\rm{GTO}}_n(\bm{r}; \bm{R_m}, \alpha_n, \mathpzc{i}_m, \mathpzc{j}_m,\mathpzc{k}_m).
    \end{aligned}
    \label{eq:CGTO}
\end{equation}
$d_n$ are called ``contraction coefficients'' as multiple concentric GTOs with equal angular momentum and different weights are ``contracted'' to approximate a Slater-type orbital (STO). STO is a type of orbital that accurately represents AOs~\cite{szabo2012modern}. The function curve of STO has a cusp near the orbital center due to its non-vanishing gradient, which GTO is incapable of reconstructing. However, forming a CGTO with appropriate contraction coefficients instead of a single GTO can alleviate this problem while retaining the computational efficiency of GTO~\cite{stewart1970small}. In our basis set generation framework, a basis function does not need to be a CGTO but can also be the linear combination of GTOs that may not have the same center position or angular momentum. We define this type of generalized Gaussian-based orbital as \textbf{mixed-contracted} Gaussian-type orbital (MCGTO): 
\begin{equation}
    \phi_{m'}^{\rm{MCGTO}}(\bm{r}) \equiv \sum_m \phi_m^{\rm{CGTO}}(\bm{r}; \bm{R_m}, \mathpzc{i}_m, \mathpzc{j}_m, \mathpzc{k}_m).
    \label{eq:MCGTO}
\end{equation}
The expressiveness of MCGTO in our framework provides the freedom to construct more delocalized basis functions than a GTO to incorporate the interatomic bonding information before applying the SCF methods. For instance, one can use the molecular orbitals of small systems as the basis functions for larger combined systems.

Furthermore, primitive variables that correlate multiple CGTOs through mapping functions add another layer of customizability to basis set generation. These correlations can encode additional physical information. For example, homonuclear diatomic molecules have spatial symmetry that can be described by the point group $D_{\infty h}$, i.e., they have rotational symmetry about their internuclear axis and reflection symmetry about a plane perpendicular to the axis~\cite{bishop1993group}. Such molecular symmetry affects the formation of their molecular orbitals. Therefore, it is reasonable to correlate the parameters of AOs during the atomic basis set generation so that such symmetry is imposed to reduce the computational cost for later parameter optimization. 

\begin{figure}[htbp]
    \centering
    \includegraphics[width=0.48\textwidth]{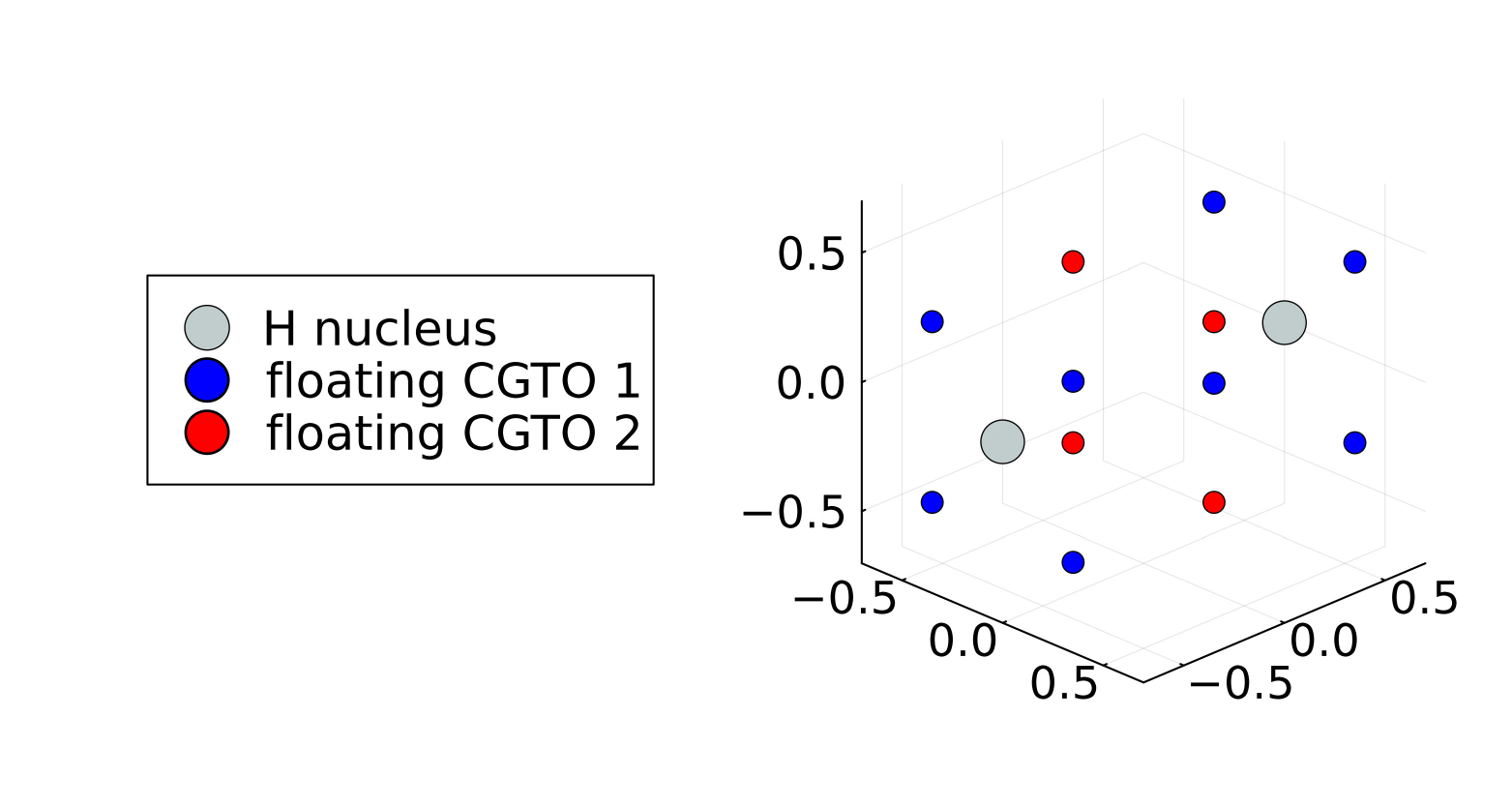}
    \caption{A $1\!\times\!2\!\times\!1$ layered grid-based Gaussian basis set for $\rm{H_2}$. Based on the geometric symmetry of $\rm{H_2}$, constraints on CGTO parameters are imposed. The CGTOs located on the grid points with the same color can have identical exponent and contraction coefficients. Correlated grid points can be directly generated using the helper function \texttt{GridBox} in Quiqbox.}
    \label{fig:gb}
\end{figure}

By using primitive variables to correlate basis functions, symmetries irrelevant to the specific molecular systems can also be applied to a basis set. For example, in a grid-based Gaussian basis set, the center position of each basis function would also be the location of a grid point controlled by the spacing of the grid points. A mapping function from the single spacing parameter (i.e., the sparsity of gird points) to the floating CGTO center positions greatly lowers the total number of basis set parameters while maintaining the topology of the grid box. The exponent and contraction coefficients of the floating CGTOs can also be correlated to reduce parameters further. In FIG.~\ref{fig:gb}, we show an example of applying both the point-group symmetry of H$_2$ molecule and the translational symmetry of grid points to construct a basis set.

\begin{figure*}[htpb]
    \centering
    \begin{subfigure}[b]{0.65\textwidth}
         \centering
         \includegraphics[width=0.99\textwidth]{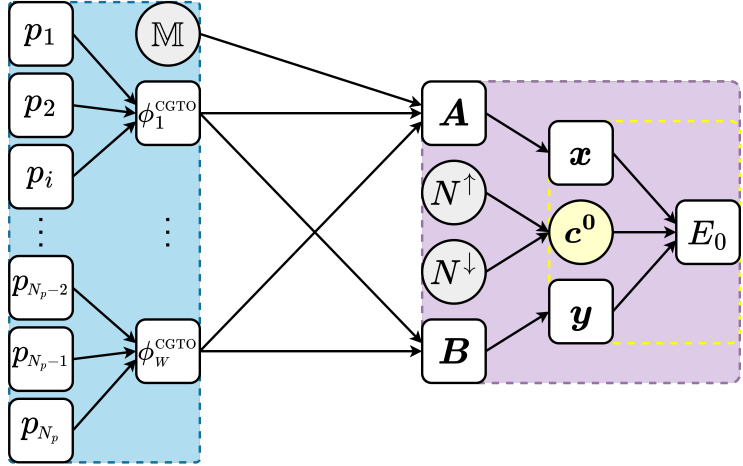}
    \caption{The complete computational graph of an algorithm solving for the electronic ground-state energy with current standard basis set generation method.}
         \label{fig:NN1}
     \end{subfigure}
     \hfill
     \begin{subfigure}[b]{0.65\textwidth}
         \centering
         \includegraphics[width=0.99\textwidth]{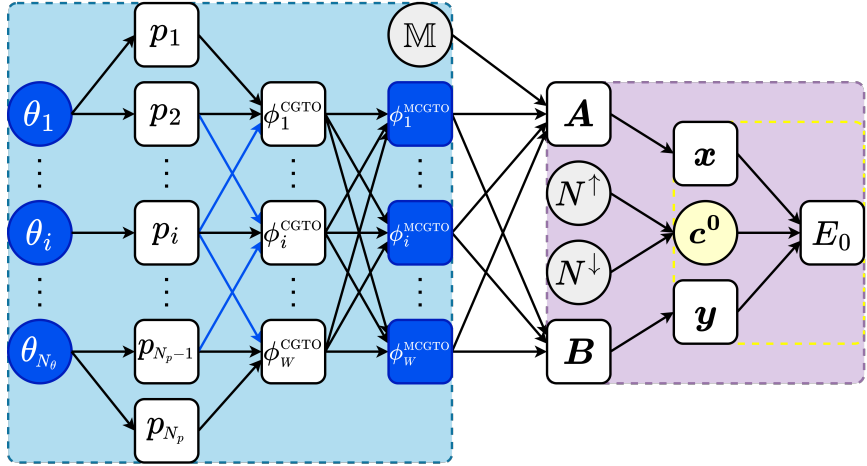}
         \caption{The computational graph with extra connectivity and layers after replacing the basis set generation method in FIG. \ref{fig:NN1} with our framework. In the light blue box, the leftmost added layer represents the primitive parameters $\{\theta_i\}$ that control CGTO parameters $\{p_i\}$ through mapping functions, which can be either linear or nonlinear functions. The rightmost added layer is composed of MCGTOs as the output nodes (basis functions). Since the CGTOs can share parameters in Quiqbox, more connectivity (blue lines) between the CGTO parameters and CGTOs also becomes available.}
         \label{fig:NN2}
     \end{subfigure}
    \caption{The comparison between the current standard basis set generation method and our framework in finding the electronic ground-state energy. The light blue box represents the basis set construction. The light purple box represents the electronic structure algorithm to compute the approximated electronic ground-state energy $E_0$. The yellow dashed box represents the chosen eigensolver that produces $\bm{c^o}$ as the final input to obtain $E_0$.}
    \label{fig:EANN}
\end{figure*}

Combining the generalized Gaussian-based orbital, MCGTO, and the additional correlation configuration among orbital parameters, we form a more customizable generation method of Gaussian-based basis sets. To provide a straightforward comparison to the currently common generation method, the computational graphs of both methods in the context of electronic ground-state energy computation are shown in FIG.~\ref{fig:EANN}. 

Aside from the system-specific parameters indicated in Eq.~(\ref{eq:Helec}), the parameters of the generated basis set $\{\phi_n\}$ also affect $E_0$. In particular, for the typical Gaussian basis set generation method (FIG.~\ref{fig:NN1}), the CGTO parameters $\{p_i\,|\,i=1\dots N_p\}$ are such parameters. However, our basis set generation framework (FIG.~\ref{fig:NN2}) further allows more primitive parameters $\{\theta_i\,|\,i=1\dots N_\theta\}$ to control these orbital parameters. On top of this, correlations between CGTOs can be implemented with shared orbital parameters among them, and MCGTOs serve as the finally generated basis functions. As we fix the coefficients generated by the eigensolver as shown in the yellow dashed box (along with the nuclei information $\mathbb{M}$ in the grey circled boxes) in FIG.~\ref{fig:EANN}, $E_0$ is mapped from $\{\theta_i\}$ through layers of differentiable nodes connected by various mapping functions. Thus, $\{\theta_i\}$ as the basis set parameters can be optimized using variational methods based on the analytical gradient of $E_0$ with respect to them, which forms the process of basis set optimization.

\section{Variational optimization of basis set parameters} \label{sec:optPars}

\subsection{Generalized analytical electronic structure gradient with respect to basis set parameters}\label{sec:generaldE}

Let $\{\theta_i\,|\,\theta_i\!\in\!\mathbb{R},\, i\!=\!1,\dots, N_\theta\}$ be a set of unique real parameters (meaning they are distinguishable from each other by their symbols and indices in spite of whether they have the same value) that $E_0$ explicitly or implicitly depends on. To minimize $E_0$ with respect to $\{\theta_i\}$, we need to compute the gradient of $E_0$ with respect to the vectorized parameters $\bm{\theta}\!=\![\theta_1, \dots, \theta_i, \dots, \theta_{N_\theta}]^T$, the analytical form of which can be expressed as~\cite{jo1983ab, hurley1988computation}: 
\begin{equation}
    \frac{dE_0}{d\bm{\theta}} = \sum_m\frac{\partial E_0}{\partial \varphi_m}\frac{\partial \varphi_m}{\partial \bm{\theta}} + \braket{\Psi_0|\frac{\partial H_{\rm elec}}{\partial \bm{\theta}}|\Psi_0}.
    \label{eq:dE0}
\end{equation}
The first term in Eq.~(\ref{eq:dE0}) represents the explicit dependency of the parameters based on the ground state approximation, and the second term represents the effective force from the Hellmann--Feynman theorem~\cite{hurley1988computation, politzer2018hellmann}. For the purpose of a systematic basis set optimization, $\{\theta_i\}$ represents a set of unique and independent basis set parameters to be optimized. Assuming the electronic Hamiltonian is independent of these parameters (e.g., the basis function centers can be detached from the nuclei locations, forming floating basis functions~\cite{hurley1988computation, frost1967floating, huber1979geometry}), the second term in Eq.~(\ref{eq:dE0}) is eliminated. Thus, the partial derivative of $E_0$ with respect to a single basis set parameter $\theta_i$ reduces to
\begin{equation}
    \frac{\partial E_0}{\partial \theta_i} = \sum_m\frac{\partial E_0}{\partial \varphi_m}\frac{\partial \varphi_m}{\partial \theta_i}.
    \label{eq:E0}
\end{equation}

Combining Eq.~(\ref{eq:E0}) with Eqs.~(\ref{eq:E0general}--\ref{eq:B}), we can derive a more detailed formula of the parameter partial derivative for a general electronic structure algorithm:
\begin{equation}
    \begin{aligned}
        \frac{\partial E_0}{\partial \theta_i}\!=&\!\sum_j\frac{\partial E_0}{\partial x_j}\frac{\partial x_j}{\partial \bm{A}}\hspace{-0.1em}\cdot\hspace{-0.1em}\frac{\partial \bm{A}}{\partial \theta_i}\!+\!\sum_k\frac{\partial E_0}{\partial y_k}\frac{\partial y_k}{\partial \bm{B}}\hspace{-0.1em}\cdot\hspace{-0.1em}\frac{\partial \bm{B}}{\partial \theta_i}\\ 
        =&\!\sum_{mn}\hspace{-0.1em}(\hspace{-0.1em}\nabla_{\hspace{-0.3em}\bm{A}}\hspace{-0.1em}E_0 \hspace{-0.25em}\odot\hspace{-0.25em} \partial_{\theta_i}\hspace{-0.1em}\bm{A}\hspace{-0.1em})_{\!mn} \hspace{-0.25em}+\hspace{-0.55em} \sum_{mnpq}\hspace{-0.2em}(\hspace{-0.1em}\nabla_{\hspace{-0.3em}\bm{B}}\hspace{-0.1em}E_0 \hspace{-0.25em}\odot\hspace{-0.25em} \partial_{\theta_i}\hspace{-0.1em}\bm{B}\hspace{-0.1em})_{\!mnpq} \label{eq:dE1}
    \end{aligned}
\end{equation}
where $\odot$ denotes element-wise multiplication, and
\begin{align}
    \begin{split}
        \hspace{-1.2em}(\nabla_{\bm{A}}E_0)_{mn} =&\, (\sum_j\frac{\partial E_0}{\partial x_j}\frac{\partial x_j}{\partial \bm{A}})_{mn},\label{eq:dEdA}
    \end{split}\\
    \begin{split}
        \hspace{-1.2em}(\nabla_{\bm{B}}E_0)_{mnpq} =&\, (\sum_k\frac{\partial E_0}{\partial y_k}\frac{\partial y_k}{\partial\bm{B}})_{mnpq},\label{eq:dEdB}    
    \end{split}\\
    \begin{split}
        (\partial_{\theta_i} \bm{A})_{mn} =&\, (\partial_{\theta_i} \varphi_m\hspace{0.1em}|\hspace{0.1em}\varphi_n) + (\varphi_m\hspace{0.1em}|\hspace{0.1em}\partial_{\theta_i} \varphi_n),\label{eq:dA}    
    \end{split}\\
    \begin{split}
        (\partial_{\theta_i} \bm{B})_{mnpq} =&\, 
        (\partial_{\theta_i} \varphi_m\varphi_n\hspace{0.1em}|\hspace{0.1em}\varphi_p\varphi_q)\hspace{0.1em} + \hspace{0.1em}\\&\,
        (\varphi_m\partial_{\theta_i} \varphi_n\hspace{0.1em}|\hspace{0.1em}\varphi_p\varphi_q)\hspace{0.1em} + \hspace{0.1em}\\&\,
        (\varphi_m\varphi_n\hspace{0.1em}|\hspace{0.1em}\partial_{\theta_i} \varphi_p\varphi_q)\hspace{0.1em} + \hspace{0.1em}\\&\,
        (\varphi_m\varphi_n\hspace{0.1em}|\hspace{0.1em}\varphi_p\partial_{\theta_i} \varphi_q).
        \label{eq:dB}
    \end{split}
\end{align}
The term $\sum_l\frac{\partial E_0}{\partial c_l}\frac{\partial c_l}{\partial \theta_i}$ is not included in Eq.~(\ref{eq:dE1}) because $\nabla_{\bm{c}}E_0$ is always a null vector at $\bm{c}=\bm{c^o}$. This is under the assumption that $\tilde{E_0}(\bm{c})$ is computed variationally, and $E_0$ is the global minimum of $\tilde{E_0}(\bm{c})$ at $\bm{c^o}$, cf. Eqs.~(\ref{eq:E0v}) and (\ref{eq:minimizeE}). Therefore, it should be noted that Eq.~(\ref{eq:dE1}) no longer holds true if $E_0$ is obtained through non-variational methods such as traditional coupled cluster method~\cite{bartlett2007coupled, helgaker2013molecular, anand2022quantum} and M\o lle--Plesset perturbation theory~\cite{helgaker2013molecular, cremer2011moller}.

To further expand Eqs.~(\ref{eq:dA}) and (\ref{eq:dB}), based on Eq.~(\ref{eq:no2o}), the partial derivatives of the orthonormal $\{\varphi_m\}$ can be expressed as~\cite{jo1983ab}:
\begin{equation}
    \begin{aligned}
        \partial_{\theta_i}\varphi_m \hspace{-0.1em}&=\hspace{-0.1em}\sum_{n} X_{nm}\partial_{\theta_i}\phi_n + \phi_n (\partial_{\theta_i} X)_{nm},\\
        (\partial_{\theta_i} X)_{mn} \hspace{-0.1em}&=\hspace{-0.1em} -\!\hspace{-0.1em}\sum_{pq}\! V_{mp}\hspace{-0.1em}\frac{(V^\dagger\partial_{\theta_i} SV)_{pq}}{(\lambda_p\lambda_q)^{1/2}(\lambda_p^{1/2}+\lambda_q^{1/2})}\hspace{-0.1em}V^*_{nq}, \\
        (\partial_{\theta_i} S)_{mn} \hspace{-0.1em}&=\hspace{-0.1em} \braket{\partial_{\theta_i}\phi_m|\phi_n} + \braket{\phi_m|\partial_{\theta_i}\phi_n}, 
    \end{aligned}
    \label{eq:dE3}
\end{equation}
where $\bm{V_p} = [V_{1p},\dots, V_{Wp}]^T$ is the $p$th eigenvector of $\bm{S}$ with corresponding eigenvalue $\lambda_p$. 

We can even further expand the partial derivative of each $\phi_n$ as the product of two terms within the summation:
\begin{equation}
    \partial_{\theta_i}\phi_n =\!\! \sum_{p_j\in\{p_j^n\}}\!\!\frac{\partial\phi_n}{\partial p_j}\frac{\partial p_j}{\partial\theta_i}
    \label{eq:dphi}
\end{equation}
where $\{p_j^n\}$ are all the CGTO parameters 
% (i.e., the exponent coefficients, the contraction coefficients, and the center position)
$\phi_n$ explicitly depends on and is differentiable with respect to: 
\begin{equation}
    \begin{aligned}
        \{p_j^n\} \!=\! \{&\alpha_1^n\!,\mydots,\alpha^n_{N^n_{\rm{GTO}}}\!, d_1^n\!,\mydots,d^n_{N^n_{\rm{GTO}}}\!, 
        X^n_1\!, \\ &Y^n_1\!, Z^n_1\!, \mydots, X^n_{N^n_{\rm{cen}}}\!, Y^n_{N^n_{\rm{cen}}}\!, Z^n_{N^n_{\rm{cen}}}\}.
    \end{aligned}
\end{equation}
The superscript $n$ of the included CGTO parameters (e.g., $\alpha_1^n$) indicates that they are corresponding to $\phi_n$. Specifically, $N^n_{\rm{GTO}}$ is the total number of GTOs in $\phi_n$, and $N^n_{\rm{cen}}$ is the total number of GTO centers that group the GTOs in $\phi_n$ into concentric GTOs. Thus, all the explicit (differentiable) CGTO parameters from $\{\phi_n\}$ would be 
\begin{equation}
    \{p_j\}  = \{p_j^1\} \cup \{p_j^2\} \cup \dots \cup \{p_j^W\}.
    \label{eq:ebsp}
\end{equation}
In the recent Gaussian basis set optimization method~\cite{tamayo2018automatic}, all basis functions in 
$\{\phi_n\}$ are CGTOs, so $N^n_{\rm{cen}}=1$, and the basis set parameters are just a subset of all the explicit parameters as shown in Eq.~(\ref{eq:ebsp}). This effectively reduces Eq.~(\ref{eq:dphi}) to only one term, a partial derivative with respect to $\theta_i$, as $\partial p_j/\partial \theta_i = \delta_{ij}$.

By calculating the analytical parameter gradient, one can perform differentiation-based parameter optimization for the generated basis set. In fact, optimization of minimal atomic basis sets with respect to the HF energy has been implemented in a Python software package named ``DiffiQult''~\cite{tamayo2018automatic}. Optimizing the basis set parameters with respect to the HF energy is a good starting point for many more advanced electronic structure algorithms, such as single-reference and multi-reference configuration interaction computations that are based on the HF state. Thus, when we try to implement the generalized basis set optimization in Quiqbox, we also set the objective function of basis set parameter optimization to the HF energy. Specifically, in the case of RHF, Eq.~(\ref{eq:E0general}) would be specialized as~\cite{szabo2012modern}
\begin{equation}
    \begin{aligned}
        \bm{T^x} =& f_1(\bm{A}) := \bm{A},\\
        \bm{T^y} =& f_2(\bm{B}) := \bm{B},\\
         \tilde{E_0}(\bm{D}) =&  f_3(\bm{D};\,\bm{T^x},\,\bm{T^y})\\ :=&\sum_{\mu,\nu}\hspace{-0.2em}D_{\!\nu\mu}({T^{\bm{x}}}_{\hspace{-0.8em}\mu\nu}\hspace{-0.2em}+\hspace{-0.3em}\sum_{\lambda,\sigma}\hspace{-0.2em}D_{\!\lambda\sigma}[2{T^{\bm{y}}}_{\hspace{-0.7em}\mu\nu\sigma\lambda}\!-\!{T^{\bm{y}}}_{\hspace{-0.7em}\mu\lambda\sigma\nu}]),
    \end{aligned}
    \label{eq:dHF}
\end{equation} 
where $\bm{D}$ is the charge density matrix of the electrons with the same spin, which can be treated as a matrix format of $\bm{c}$. Similarly, $\bm{T^x}$ and $\bm{T^y}$ are, respectively, $\bm{x}$ and $\bm{y}$ in tensor formats.

\subsection{Variational parameter optimization based on hybrid differentiation design}

If one can find all the elementary differentiation rules required to compute Eq.~(\ref{eq:dE1}), along with necessary rules for related program subroutines, in a reliable AD software library, they will be able to finish the code for the differentiation-based optimization with minimal effort. Specifically, reverse-mode AD would be the preferable AD mode to implement as the number of basis set parameters is normally larger than the output dimension of the objective function (e.g., $E_0$ for the HF energy)~\cite{baydin2018automatic}. However, reverse-mode AD is not well defined when the diagonalization process of matrices with degenerate eigenvalues emerges in the programming code. This led to the compromise of using the rather inefficient forward-mode AD to perform the gradient computation in DiffiQult~\cite{tamayo2018automatic}.

Instead of fully relying on forward-mode AD or adapting reverse-mode AD with customized AD rules as a workaround, we designed a hybrid differentiation engine in Quiqbox. This differentiation engine combines both AD and symbolic differentiation (SD) so that it can provide both efficiency and extensibility to Quiqbox's parameter optimization procedure. The fundamental design of it is to divide the parameter differentiation, i.e., Eq.~(\ref{eq:dE1}), into three parts:
\begin{enumerate}
    \item compute $\nabla_{\bm{A}}E_0$ and $\nabla_{\bm{B}}E_0$;
    \item compute $\partial_{\theta_i} \bm{A}$ and $\partial_{\theta_i} \bm{B}$;
    \item compute \\$\partial_{\theta_i} E_0\!=\!f_4(\nabla_{\!\hspace{-0.2em}\bm{A}}\hspace{-0.1em}E_0,\nabla_{\!\hspace{-0.2em}\bm{B}}\hspace{-0.1em}E_0, \partial_{\theta_i} \bm{A},\partial_{\theta_i} \bm{B})$ where
\end{enumerate}
\begin{equation}
    \begin{aligned}
        &f_4(\bm{T^a},\:\bm{T^b},\:\bm{T^c},\:\bm{T^d})\\
        :=& \sum_{mn} (\bm{T^a}\!\odot\!\bm{T^c})_{mn}\!+\!\sum_{mnpq}\hspace{-0.2em}(\bm{T^b}\!\odot\!\bm{T^d})_{mnpq}.
    \end{aligned}
    \label{eq:contraction}
\end{equation}
The first part requires the computation of Eqs.~(\ref{eq:dEdA}) and (\ref{eq:dEdB}), hence it only depends on the choice of electronic structure algorithm. The second part requires the computation of Eqs.~(\ref{eq:dA}) and (\ref{eq:dB}), and thus it only depends on the generated basis set. The third part is simply tensor contractions. 

By dividing the chain of analytical differentiation into computational modules, we lose the capability of utilizing a single full AD computation. However, since each module can be independently maintained and updated, doing so can reduce the risk of code breaking due to bugs coming from external AD libraries. This is crucial from the perspective of developing sustainable scientific programs. More importantly, we now gain the flexibility to design tailored optimization for each module.

Particularly, for part one, one can choose between AD and SD depending on the complexity of Eqs.~(\ref{eq:dEdA}) and (\ref{eq:dEdB}) determined by the applied electronic structure algorithm. As for part two, the differentiation of each $\phi_n$ can be optimized by transforming it into the efficient Gaussian-based electronic integral computation. Due to the generality of MCGTO, it can represent any linear combination based on Eqs.~(\ref{eq:GTO}--\ref{eq:MCGTO}). Therefore, any partial derivative of MCGTO at a fixed point, including a CGTO, can be fully expressed by a new MCGTO. This means the expression of MCGTO is complete for arbitrary order of derivatives of MCGTO. Hence, we can define a new basis function $\psi_n^{\theta_i}$ with respect to $\theta_i$ and $\phi_n$ such that
\begin{equation}
    \psi_n^{\theta_i} \equiv \partial_{\theta_i}\phi_n.
\end{equation}
This effectively transforms the computation of Eqs.~(\ref{eq:dA}) and (\ref{eq:dB}) into the computation of electronic integrals of MCGTOs that compose a new \textbf{derivative basis set}:
\begin{equation}
    \mathbb{D} \equiv \{\phi_n\} \cup \{\psi_n^{\theta_1}\} \cup \dots \cup \{\psi_n^{\theta_{N_\theta}}\}.
\end{equation}
Quiqbox utilizes AD to compute all the partial derivatives that directly contribute to contraction coefficients and uses SD to generate new MCGTOs as the partial differentiation of basis functions. Due to the multiple dispatch feature and the just-in-time (JIT) compilation in the Julia programming language~\cite{bezanson2018julia}, $\mathbb{D}$ can be generated automatically and efficiently. In the current version (0.5.7), Quiqbox computes and stores all the intermediate components of the gradients ($\nabla_{\bm{A}}E_0$, $\nabla_{\bm{B}}E_0$, $\partial_{\theta_i} \bm{A}$, and $\partial_{\theta_i} \bm{B}$) before performing the tensor contraction, cf. Eq.~(\ref{eq:contraction}), at the end of the differentiation process. Consequently, the space complexity (memory cost) of such a procedure has a lower bound of $\Omega(W^4)$ and an upper bound of $\mathcal{O}(N_\theta W^4)$ depending on whether the partial derivatives are computed sequentially or in embarrassingly parallel.

\begin{figure}[ht]
    \centering
    \includegraphics[width=0.99\linewidth]{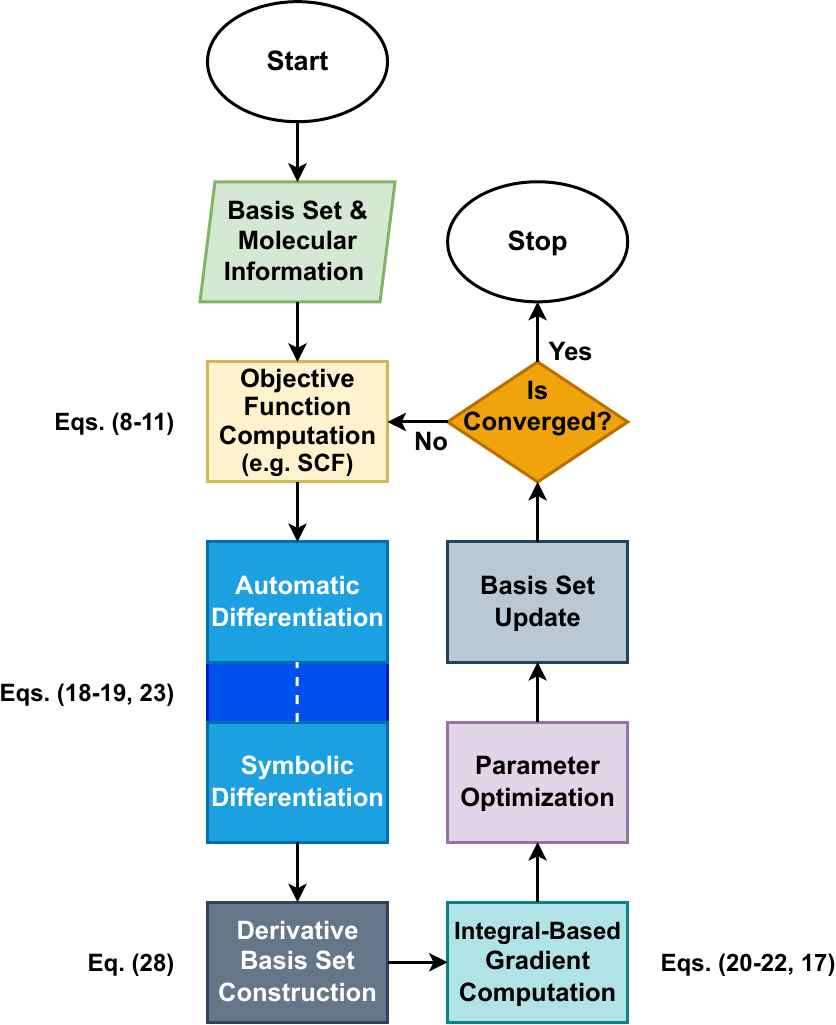}
    \caption{The diagram of the basis set optimization procedure of Quiqbox. Within each iteration cycle of the optimization, both automatic differentiation (AD) and symbolic differentiation (SD) are applied to compute the parameter gradient. Particularly, partial derivatives of basis functions with respect to the parameters are then transformed into a new derivative basis set $\mathbb{D}$ such that part of gradient computation can be transformed into a direct evaluation of electronic integrals (shown in the teal box).}
    \label{fig:po}
\end{figure}

The architecture of Quiqbox's parameter optimization procedure applied with the aforementioned hybrid-differentiation design is shown in FIG.~\ref{fig:po}. The computation of the objective function is isolated from the computation of the gradient, as the former provides a stationary point of $\bm{c}$ such that $\tilde{E_0}(\bm{c}) \approx E_0$ in order to proceed the latter. As a result, the option of objective functions can also be extended separately by programming new computational modules. This allows Quiqbox to adapt variational algorithms more advanced than HF without overhauling the overall workflow of basis set optimization. 

In the following three sections, we give examples of utilizing Quiqbox for various basis set optimization and generation applications. Unless specifically stated otherwise, all the data are computed solely using Quiqbox. The gradient-based optimizer used by Quiqbox is L-BFGS~\cite{liu1989limited}. For Quiqbox's performance (as of version 0.5.7) for basis set optimization and ab initio computation, please refer to Appx.~\ref{app:performance}.

\section{Example I: Optimizing atomic basis sets on the Hartree--Fock level} \label{sec:AbsOpt}

One immediate application of Quiqbox is to optimize existing atomic basis sets. Given that the currently implemented electronic structure algorithms in Quiqbox are HF methods, we shall demonstrate some results of optimizing the CGTO parameters ($\alpha$, $d$, and $\bm{R}$ unless specified otherwise) of basis sets for small molecules on the level of the Hartree--Fock approximation.

\subsection{Dissociation curves of $\rm H_2$ and $\rm Li_2$}\label{sec:H2Li2}

In quantum chemistry, the potential energy surface (PES) with respect to the dissociation of a molecular system into smaller constituents is just as important, if not more important, than the ground-state energy of the system at its equilibrium geometry. Suppose the optimized basis set provides no improvement on the overall shape of the PES curve but only a constant shift compared to the original basis set. In that case, the practicality of optimizing basis sets based on energy gradient will diminish, as the optimized basis set does not provide a better description of energy change during the dissociation process than the original one.

\begin{figure*}[htbp]
    \centering
    \begin{subfigure}[b]{0.48\textwidth}
         \centering
         \includegraphics[width=0.99\textwidth]{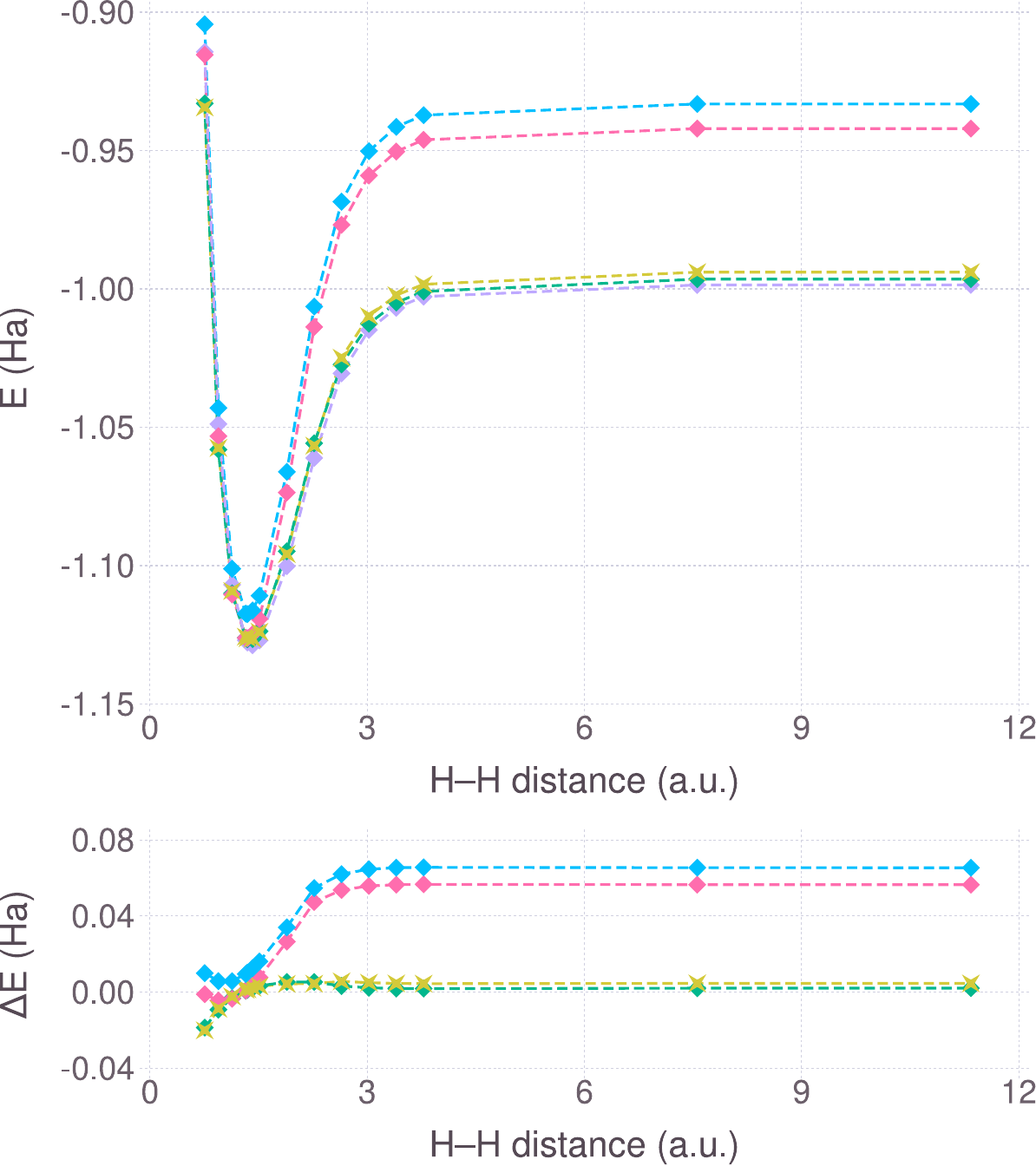}
         \caption{Dissociation curves of $\rm H_2$.}
         \label{fig:PES_H2}
     \end{subfigure}
     \hspace{0.5em}
     \begin{subfigure}[b]{0.48\textwidth}
         \centering
         \includegraphics[width=0.99\textwidth]{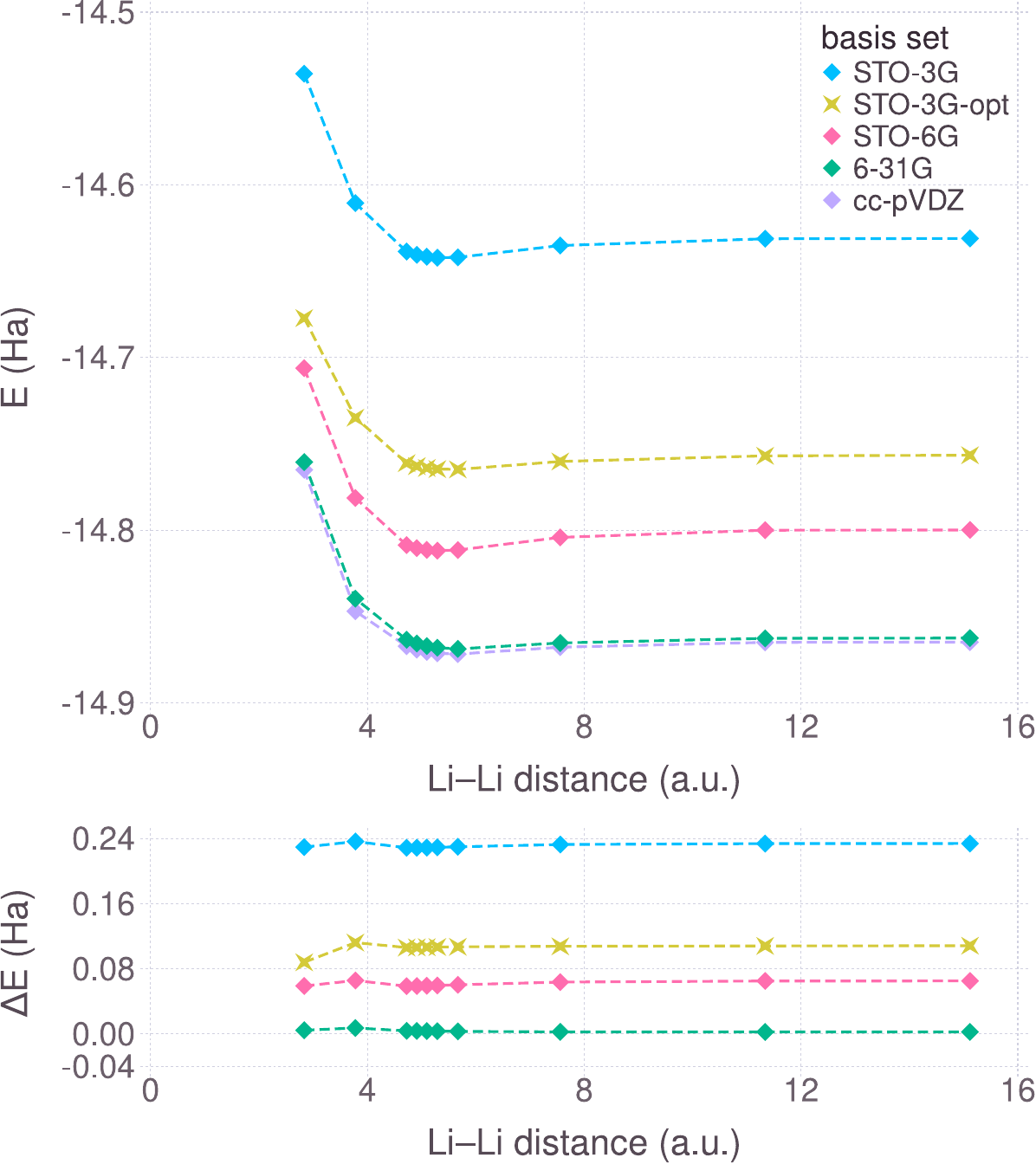}
         \caption{Dissociation curves of $\rm Li_2$.}
         \label{fig:PES_Li2}
     \end{subfigure}
    \caption{The unrestricted open-shell Hartree--Fock (UHF) potential energy surfaces with respect to the bond length in the cases of $\rm H_2$ and $\rm Li_2$. The label ``STO-3G-opt'' in each sub-figure represents the optimized STO-3G basis set for the corresponding molecules. The surfaces of the HF energy with the nuclear repulsion included ($\rm E$) are plotted in the upper-larger canvas of each sub-figure. The energy differences ($\Delta \rm E$) between using cc-pVDZ and each one of the other four basis sets are shown in the lower-flat canvas of each sub-figure.}
    \label{fig:PES}
\end{figure*}

To show that the boost of the Hartree--Fock approximation from optimized atomic basis sets can be non-trivial, we plot the dissociation curves of optimized STO-3G basis sets in the cases of $\rm H_2$ and $\rm Li_2$ as shown in FIG.~\ref{fig:PES}. Specifically, in FIG.~\ref{fig:PES_H2}, the optimized STO-3G basis set (labeled as STO-3G-opt) provides both a lower energy globally and a better energy surface very close to the higher-quality basis sets 6-31G~\cite{hehre1972self} and cc-pVDZ~\cite{dunning1989gaussian}. On the contrary, the STO-6G basis set, which requires more GTOs to build, is mainly a constant shift of STO-3G. As a result, the dissociation energy (by the supermolecule approach~\cite{helgaker2013molecular}) of $\rm H_2$ in STO-3G-opt ($\rm 0.132 Ha$) is much closer to its value computed in cc-pVDZ ($\rm 0.130 Ha$), as supposed to the results in STO-3G and STO-6G (both of which are $\rm 0.184 Ha$).

In the case of $\rm Li_2$, the optimized STO-3G basis set (labeled as STO-3G-opt as well in FIG.~\ref{fig:PES_Li2}) no longer recovers a high percentage of the HF energy in cc-pVDZ or 6-31G compared to the case of $\rm H_2$. However, it still provides a better agreement to cc-pVDZ than both STO-3G and STO-6G from the region of equilibrium geometry  (around Li--Li distance equal to 5 a.u.) to the bond-breaking region (where Li--Li distance is beyond 10 a.u.). This is shown more clearly in the lower half figure of FIG.~\ref{fig:PES_Li2}.

Based on the above two examples, we have shown that we can improve the performance of a basis set for generating a molecular system's dissociation curve without adding more basis functions. Specifically, optimizing the GTO parameters with respect to the ground-state ansatz, which in this case is the HF state, can improve the performance. However, given a fixed number of primitive GTOs in each basis function, the improvement it can give is specific to the target system (STO-3G-opt in $\rm Li_2$ provides less improvement than in $\rm H_2$). Thus, it would be of interest to quantitatively study how the number of GTOs used to construct CGTOs as basis functions will affect the performance of an atomic basis set for molecular systems. Due to Quiqbox's flexible basis set generation and optimization framework, it is already well-equipped for this potential future research direction.

\subsection{Atomizations and formations involving hydrogen and oxygen}\label{sec:H2O2}

Aside from the systematic error of the ground-state energy across different geometries, the transferability of optimized atomic basis sets across different molecules is also crucial~\cite{Silver1978}. We refer to ``transferability'' as the property of basis sets to provide consistent performance of calculating the same physical observable for the different molecular systems. For atomic basis sets, the basis sets from the same family (e.g. 6-31G) are usually constructed using the same protocol. By using atomic basis sets from the same family to compute properties like atomization energy and chemical reaction enthalpy, some systematic errors generated by the basis sets can be canceled out when taking the energy differences~\cite{bak2000accuracy, jensen2013atomic}. Since MCGTO based on optimized atomic basis sets are, by default, generated molecule by molecule as opposed to directly grouping AOs, it is reasonable to question their transferability compared to the original basis sets. 

%6-311G//STO-3G

Conducting a comprehensive study of the transferability of atomic basis sets and their optimized version would require additional survey and analysis that deviates from the scope of this paper. Thus, in light of this potential research direction, we only provide a minimal test example where we compute a handful of atomization and formation energies involving molecules composed of H or O.

We optimized two basis sets, STO-3G and 6-31G~\cite{hehre1972self}, with respect to the ground-state HF energy of various atomic or molecular systems at fixed molecular geometry provided by the CCCBDB~\cite{cccbdb} database (Table~\ref{tb:geoinfo} in Appx.~\ref{app:reaction} contains the specific nuclear coordinates). Then, we calculated the enthalpies of various atomization and formation reactions using the optimized basis sets (STO-3G-opt and 6-31G-opt) and reference basis sets. The results are shown in Table~\ref{tb:reactions}. For STO-3G-opt, despite providing lower energies (as shown in Table~\ref{tb:groundstate} in Appx.~\ref{app:reaction}), the corresponding reaction enthalpies are more often less accurate compared to STO-3G according to the results at extrapolated complete basis set (CBS) limit~\cite{HALKIER1999437}. For the larger 6-31G, 6-31G-opt does provide improvements in most cases. We believe this suggests that optimizing a larger basis set with more basis functions of the same angular momentum (i.e., 6-31G does not add basis functions beyond $s$ and $p$ subshells in the cases of H and O) is more effective. 

At the same time, missing orbitals with higher angular momentum might be another limiting factor of the optimization's performance aside from inadequate basis functions. For example, due to the lack of $d$-type orbitals compared to cc-pVDZ, both 6-31G and 6-31G-opt provide a qualitatively wrong atomization energy for O$_2$. To observe the contribution of higher-angular-momentum basis functions, we constructed and tested hybrid basis sets STO-3G** and STO-3G-opt**. They are augmented versions of STO-3G and STO-3G-opt, respectively, by including additional polarization functions (orbitals) that are appended to 6-31G to construct 6-31G**~\cite{Hariharan1973}. Specifically, a set of basis functions corresponding to a full $p$-subshell is added to the basis set for H, and a set of basis functions corresponding to a full $d$-subshell is added to the basis set for O. In this way, the added basis functions are fixed across all cases. It is worth mentioning that such basis sets can be easily built using Quiqbox, as it supports grouping arbitrary basis functions into a basis set. The results of STO-3G** and STO-3G-opt** are also shown in Table~\ref{tb:reactions}, where STO-3G-opt** outperforms both STO-3G and STO-3G**. Further investigation is needed to verify if such a strategy can also provide significant improvement for other basis sets. Moreover, it would be of interest to see whether optimization on the added basis functions with higher angular momentum will further improve the performance.

In spite of the preliminary results, the tests in Sec.~\ref{sec:AbsOpt} should be mainly treated as demonstrations of potential perspectives for a systematic study of basis set optimization with the aid of Quiqbox. More tests are required to provide more insights and draw conclusions regarding the practicality and systematic behaviors of optimized atomic basis sets. We also carry out additional discussions regarding related future research directions in Sec.~\ref{sec:conclusion}.

\begin{center}
    \begin{table*}[ht]
    \resizebox{\textwidth}{!}{%
    \setlength{\tabcolsep}{2.4pt}
    \renewcommand{\arraystretch}{1.2}
    \centering
    \begin{tabular}{lrrrrrrrrrrrrrr}
        \toprule[1.5pt]
        \multirow{2}{*}{Basis Set} & \multicolumn{2}{c}{${\rm H}_2\hspace{-0.3em}\to\hspace{-0.3em}{\rm H}\hspace{-0.25em}+\hspace{-0.25em}{\rm H}$} & \multicolumn{2}{c}{${\rm O}_2\hspace{-0.3em}\to\hspace{-0.3em}{\rm O}\hspace{-0.25em}+\hspace{-0.25em}{\rm O}$} & \multicolumn{2}{c}{${\rm H}_2{\rm O}\hspace{-0.3em}\to\hspace{-0.3em}{\rm 2H}\hspace{-0.25em}+\hspace{-0.25em}{\rm O}$} & \multicolumn{2}{c}{${\rm H}_2{\rm O}_2\hspace{-0.3em}\to\hspace{-0.3em}{\rm 2H}\hspace{-0.25em}+\hspace{-0.25em}{\rm 2O}$} & \multicolumn{2}{c}{${\rm H}_2\hspace{-0.25em}+\hspace{-0.25em}\frac{1}{2}{\rm O}_2\hspace{-0.3em}\to\hspace{-0.3em}{\rm H}_2{\rm O}$} & \multicolumn{2}{c}{${\rm H}_2\hspace{-0.25em}+\hspace{-0.25em}{\rm O}_2\hspace{-0.3em}\to\hspace{-0.3em}{\rm H}_2{\rm O}_2$} & \multicolumn{2}{c}{${\rm H}_2{\rm O}\hspace{-0.25em}+\hspace{-0.25em}\frac{1}{2}{\rm O}_2\hspace{-0.3em}\to\hspace{-0.3em}{\rm H}_2{\rm O}_2$} \\ \cline{2-15} 
         & $\Delta\hspace{-0.08em}H\hspace{-0.05em}(\hspace{-0.08em}\frac{\rm kcal}{\rm mol}\hspace{-0.08em})$ & Error & $\Delta\hspace{-0.08em}H\hspace{-0.05em}(\hspace{-0.08em}\frac{\rm kcal}{\rm mol}\hspace{-0.08em})$ & Error & $\Delta\hspace{-0.08em}H\hspace{-0.05em}(\hspace{-0.08em}\frac{\rm kcal}{\rm mol}\hspace{-0.08em})$ & Error & $\Delta\hspace{-0.08em}H\hspace{-0.05em}(\hspace{-0.08em}\frac{\rm kcal}{\rm mol}\hspace{-0.08em})$ & Error & $\Delta\hspace{-0.08em}H\hspace{-0.05em}(\hspace{-0.08em}\frac{\rm kcal}{\rm mol}\hspace{-0.08em})$ & Error & $\Delta\hspace{-0.08em}H\hspace{-0.05em}(\hspace{-0.08em}\frac{\rm kcal}{\rm mol}\hspace{-0.08em})$ & Error & $\Delta\hspace{-0.08em}H\hspace{-0.05em}(\hspace{-0.08em}\frac{\rm kcal}{\rm mol}\hspace{-0.08em})$ & Error \\ \midrule[1.2pt]\midrule[1.2pt]
        STO-3G       & 115.68 & 38.37\% & 17.64  & 45.01\%  & 143.44 & 6.04\%  & 140.27 & 6.46\%  & -18.94 & 64.27\% & -6.95  & 56.78\%  & 11.99 & 67.53\% \\
        STO-3G-opt   & 82.98  & 0.74\%  & -23.90 & 174.50\% & 130.52 & 14.50\% & 100.05 & 24.07\% & -59.49 & 12.22\% & -40.97 & 154.79\% & 18.52 & 49.85\% \\
        STO-3G**     & 117.99 & 41.14\% & 54.59  & 70.17\%  & 165.23 & 8.23\%  & 176.61 & 34.04\% & -19.95 & 62.37\% & -4.03  & 74.94\%  & 15.91 & 56.92\% \\
        STO-3G-opt** & 83.08  & 0.62\%  & 20.34  & 36.60\%  & 143.81 & 5.80\%  & 128.21 & 2.69\%  & -50.56 & 4.62\%  & -24.79 & 54.17\%  & 25.77 & 30.22\% \\
        6-31G        & 81.65  & 2.33\%  & -9.78  & 130.49\% & 127.37 & 16.57\% & 91.17  & 30.81\% & -50.60 & 4.55\%  & -19.30 & 20.02\%  & 31.31 & 15.22\% \\
        6-31G-opt    & 82.89  & 0.85\%  & -6.09  & 118.98\% & 133.72 & 12.41\% & 96.86	& 26.49\% & -53.87 & 1.62\%  & -20.05 & 24.69\%  & 33.82 & 8.42\% \\
        cc-pVDZ      & 81.08  & 3.01\%  & 27.10  & 15.52\%  & 145.99 & 4.37\%  & 125.00 & 5.13\%  & -51.36 & 3.11\%  & -16.82 & 4.60\%   & 34.54 & 6.47\% \\
        CBS Limit    & 83.60  & 0.00\%  & 32.08  & 0.00\%   & 152.66 & 0.00\%  & 131.76 & 0.00\%  & -53.01 & 0.00\%  & -16.08 & 0.00\%   & 36.93 & 0.00\% \\
        \bottomrule[1.5pt]
    \end{tabular}}
    \caption{The atomization and various formation energies related to H and O computed with respect to different basis sets. Particularly, the reaction enthalpies ($\Delta H$ in kcal/mol) are calculated based on the ground-state HF energies (including nuclear repulsion). The values of them and the corresponding molecular geometries are provided in Table~\ref{tb:groundstate} and Table~\ref{tb:geoinfo} respectively in Appx.~\ref{app:reaction}.}
    \label{tb:reactions}
    \end{table*}
\end{center}

\section{Example II: Preparing the initial state for ground-state VQE computation}\label{sec:VQE}

Although we have only demonstrated the possible improvements MCGTO can provide as parameter-optimized atomic basis sets with respect to the HF energy, we hope such potential remains in post-HF computation. As mentioned in Sec.~\ref{sec:intro-nisq}, developing practical NISQ computation requires more efficient utilization of the basis set due to the limitation of qubit constraints and circuit depth. If the differentiation-based optimization of the HF state also improves the performance of the derived multi-configuration state, then one can adopt basis set optimizations as the final step of state preparations for post-HF computation.

$\rm BeH_2$ has been a popular testing molecule in multiple pieces of literature proposing ground-state VQE implementations~\cite{kandala2017hardware, xia2020qubit, chivilikhin2020mog, schleich2022improving}. Thus, we chose the co-linear gaseous form with Be--H bond length at 1.3264 \AA~as a target system to demonstrate the possibility of optimizing basis sets for ground-state VQE computation of molecular systems. We optimized the exponent and contraction coefficients to arrive at an atom-centered optimized STO-3G (STO-3G-ECopt). To impose the point-group symmetry of $\rm BeH_2$ onto the basis set, we correlated the parameters of two hydrogen STO-3G basis sets so that they remained equal throughout the optimization process in Quiqbox. The code for preparing and optimizing STO-3G is shown in Listing~\ref{lst:BeH2basis}.

{\normalsize \jlinputlisting[caption={The Julia code of using Quiqbox to optimize the exponent and contraction coefficients of the STO-3G basis set for $\rm BeH_2$.}, label={lst:BeH2basis}]{BeH2opt_code.jl}}

To compare STO-3G-ECopt to STO-3G, we used the quantum computing software library Pennylane~\cite{bergholm2018pennylane}. With Pennylane, we performed a classical simulation of VQE. The VQE circuit ansatz for approximating the many-electron ground state is constructed using parameterized Givens rotation gates~\cite{arrazola2022universal}. This is equivalent to a first-order Trotter approximation of the unitary coupled cluster with singly and doubly excitations (UCCSD)~\cite{bartlett1989alternative, romero2018strategies, anand2022quantum}. The Jordan-Wigner transformation~\cite{jordan1993paulische, nielsen2005fermionic} is used to translate the electronic Hamiltonian to a qubit Hamiltonian. The parameters of the VQE trial state are iteratively optimized. At each iteration, the gradient of the energy with respect to the parameters is estimated from the expectation values of the qubit Hamiltonian. This process is repeated until convergence of the energy is reached. For more detailed discussions of VQE, we refer the reader to reference~\cite{tilly2022variational}.

\begin{figure}
    \centering
    \includegraphics[width=0.99\linewidth]{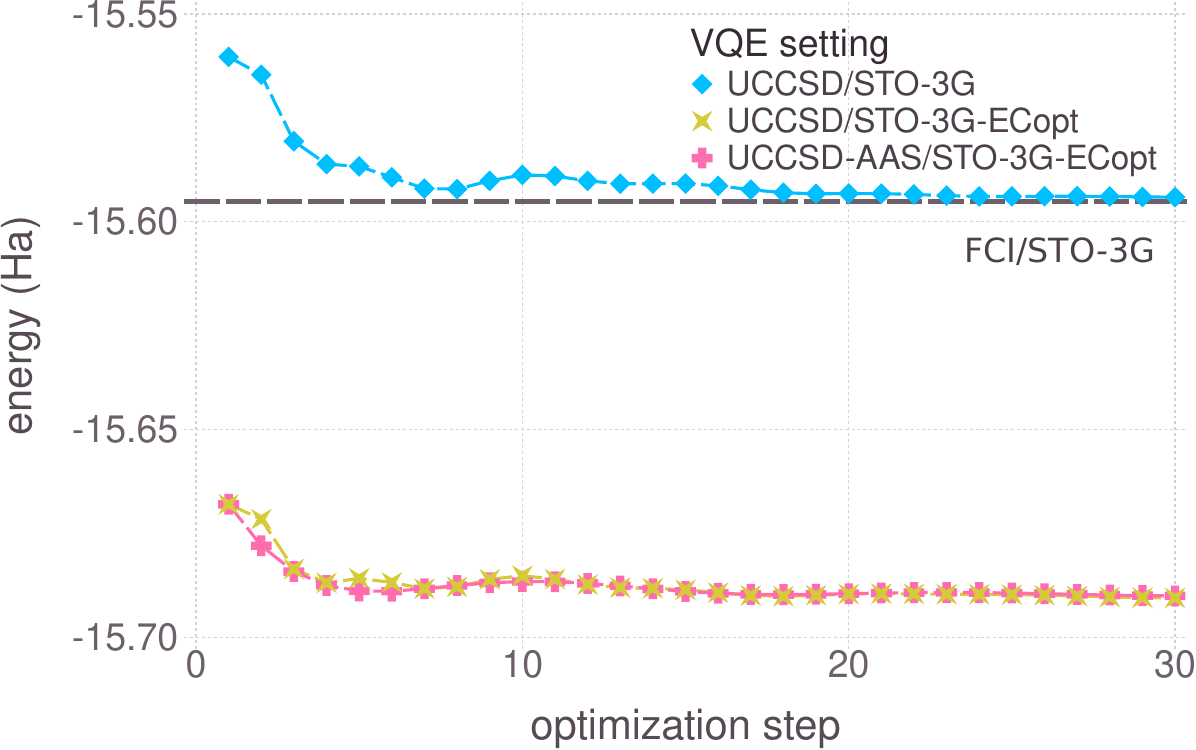}
    \caption{The optimization curves of the ground-state total energy of $\rm BeH_2$ with respect to VQE settings and basis sets. Specifically, ``STO-3G-ECopt'' refers to STO-3G with optimized exponent and contraction coefficients based on the RHF energy of $\rm BeH_2$; ``UCCSD-AAS'' refers to applying an adaptive UCCSD VQE~\cite{grimsley2019adaptive} on an active space of six molecular (spatial) orbitals with the highest energies. The vertical dashed line marks the total energy from the full configuration interaction (FCI) computation in STO-3G.}
    \label{fig:VQEcompr}
\end{figure}

The optimization curves of different VQE settings are shown in FIG.~\ref{fig:VQEcompr}. The UCCSD VQE using STO-3G-ECopt (UCCSD/STO-3G-ECopt) has an initial state with lower energy than the full configuration interaction (FCI) energy in STO-3G (shown in the horizontal dashed line). It also reaches even lower energy as it converges. We have also included the result of another test run (labeled as ``UCCSD-AAS/STO-3G-ECopt'') where an adaptive VQE~\cite{grimsley2019adaptive} with the same UCCSD ansatz was run on a selected active space of molecular orbitals of the initial HF state. We froze the molecular (spatial) orbital with the lowest energy as the one core orbital and define the active space as the rest six orbitals. The active space selection reduces the number of required qubits from 14 to 12, and the VQE adaptively adjusts the depth of the circuit during the optimization process by removing excitation operators deemed unimportant. This setting provides a more practical setting if running the VQE algorithm on an actual quantum device is desired.

Since the ansatz for the VQE is post-HF, the resulting improvement reflects the benefit of performing Hartree--Fock-level basis set optimization for multi-configuration methods. We expect this to be true for both classical and quantum implementations. 

Constructing natural orbitals~\cite{davidson1972properties} is another method to improve the convergence of CI expansion by identifying the least important orbitals which can be subsequently removed. Similarly, local and pair natural orbital schemes are used in the context of coupled cluster theory to compress the virtual space \cite{neese2009efficient, rolik2013efficient, sparta2014chemical}. Despite the multiple variants and types of natural orbitals, essentially, they are all re-optimized linear combinations of the initially given basis set. This is in contrast to the optimization procedure of our framework, which changes the initial basis set itself and cannot be replaced by linear transformations of given orbitals. In future work, it will be interesting to combine the two techniques together to determine if the natural orbital methods are equally effective with Quiqbox-optimized basis functions as with standard basis sets.

\section{Example III: Building basis sets from delocalized orbitals} \label{sec:DO}

Now we showcase the customizability MCGTO provides beyond optimized atomic basis sets. Consider a basis set $\{\phi'_n\}$, where the Coulomb interaction between every two different spatial orbitals is equal to a constant $\mathcal{J}$, and the exchange interaction between every two different spatial orbitals is equal to a constant $\mathcal{K}$, i.e., 
\begin{equation}
    \begin{aligned}
        \forall \:\phi'_i,\: \phi'_j &\in \{\phi'_n\}\: {\rm{s.t.}} \:\phi'_i \neq \phi'_j,\\ 
        \mathcal{J} &\equiv (\phi'_i\phi'_i|\phi'_j\phi'_j),\\
        \mathcal{K} &\equiv (\phi'_i\phi'_j|\phi'_j\phi'_i). 
    \end{aligned}
\end{equation}
Such basis sets can be considered to have ``completely delocalized orbitals (CDOs),'' resulting in permutation invariance of electronic modes with the same spin configuration. 

One example of such a basis set is three pairs of identical floating CGTOs that all intersect at the same point such that every line segment that connects two CGTOs (at their centers) in the same pair is a perpendicular bisector of other pairs. The two CGTOs of each pair are combined as a basis function in the basis set. If each CGTO contains $n$ GTOs, then in total, only $n$ floating GTOs with unique exponents and contraction coefficients are used to construct all three basis functions of the basis set. 

\begin{figure}[htbp]
    \centering
    \includegraphics[width=0.48\textwidth]{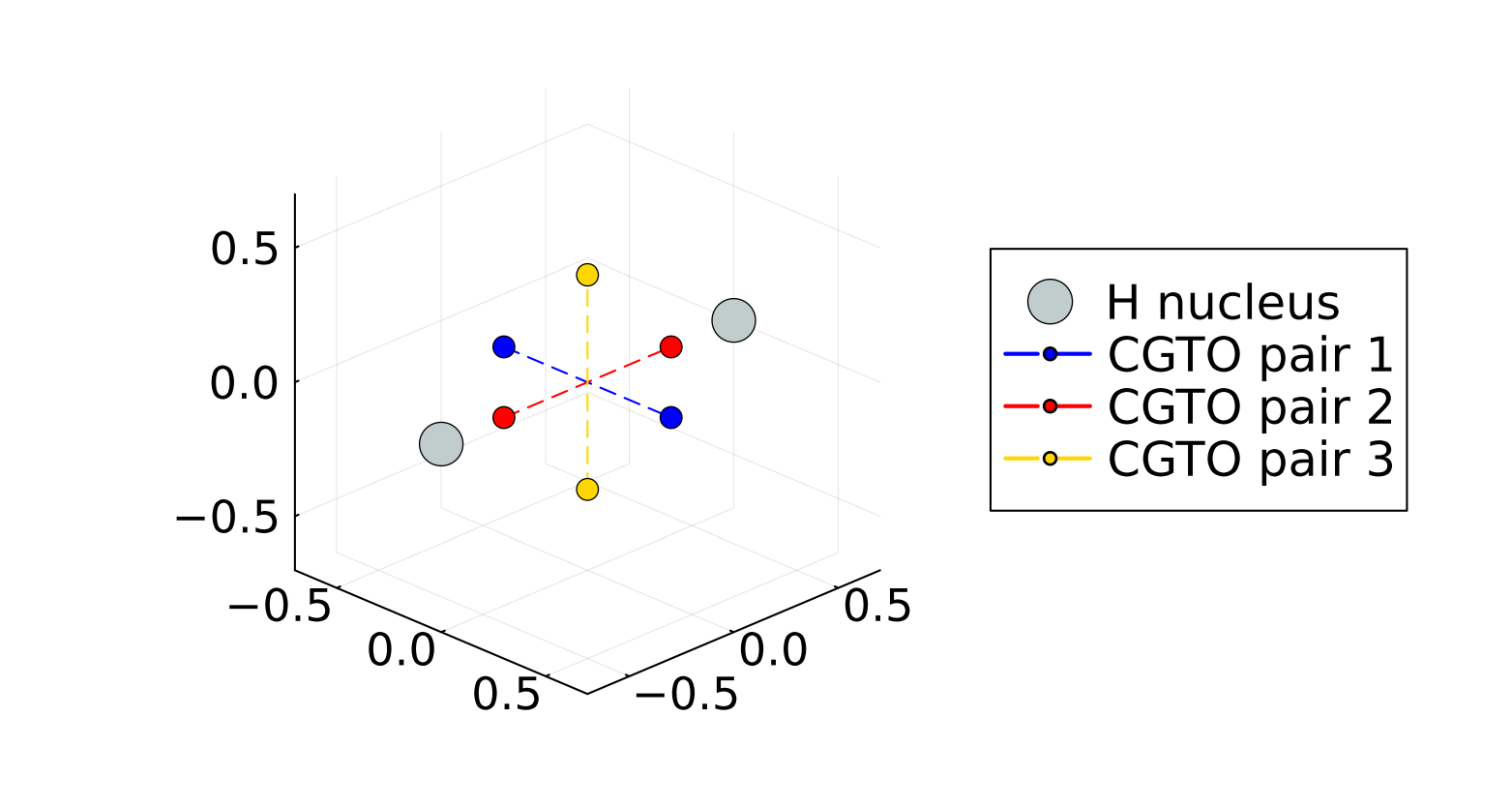}
    \caption{A basis set of three completely delocalized orbitals (CDO3) concerning $\rm H_2$. Each pair of CGTO with the same exponent and contraction coefficients, shaped like a ``dumbbell,'' forms a basis function.}
    \label{fig:CDO3}
\end{figure}

The geometry of this basis set is shown in FIG.~\ref{fig:CDO3}, and the code for constructing an instance of such a basis set with $n\!=\!4$ (i.e., CDO3-4G) in Quiqbox is shown in lines 10--31 of Listing \ref{lst:DObasis}.

{\normalsize \jlinputlisting[caption={The Julia code of generating and optimizing a CDO3-4G basis set for $\rm{H_2}$ using Quiqbox.}, label={lst:DObasis}]{DObasis_code.jl}}

According to a fermionic de Finetti theorem proven in reference~\cite{krumnow2017fermionic}, a fermionic quantum state of a finite-size system that is invariant under permutations of modes can be well described by a non-interacting fermionic state produced by mean-field methods. It would be of interest to study whether a CDO-based basis set that enforces the orbital permutation invariance would improve the performance of HF methods without losing much information from discretizing the molecular electronic Hamiltonian. We used Quiqbox to perform a parameter optimization of CDO3-4G (see lines 33--46 of Listing \ref{lst:DObasis}). The result of this optimization is compared to typical atomic basis sets in Table~\ref{tbCDOHF}. CDO3-4G is able to produce lower HF energy for $\rm{H_2}$ with one-sixth as many orbitals as aug-cc-pVDZ~\cite{kendall1992electron}. This suggests the potential advantage of CDO-based basis sets to compress the information of a fermionic ground state into a single Slater determinant, as opposed to the traditional atomic basis sets.

\begin{center}
    \begin{table}[ht]
    \setlength{\tabcolsep}{2.1pt}
    \renewcommand{\arraystretch}{1.2}
    \centering
    \begin{tabular}{lccrr}
    \toprule[1.5pt]
    Basis set   & $W$ & $N_{\rm{GTO}}$ & \multicolumn{1}{c}{$E_0\:({\rm{Ha}})$} & \begin{tabular}[c]{@{}c@{}}Error w.r.t.\\ CBS limit\end{tabular} \\ \midrule[1.2pt]\midrule[1.2pt]
    % GB-1-2-1-2G & $12$  & $24$              & $-1.76499$                            & $+4.233\%$                                                                 \\
    STO-3G      & $2$   & $6$               & $-1.83100$                            & $0.910\%$                                                                 \\
    6-31G       & $4$   & $8$               & $-1.84103$                            & $0.367\%$                                                                 \\
    cc-pVDZ     & $10$  & $16$              & $-1.84300$                            & $0.261\%$                                                                     \\
    aug-cc-pVDZ & $18$  & $24$              & $-1.84307$                            & $0.257\%$                                                                     \\
    CDO3-4G      & $3$   & $24$              & $-1.84617$                            & $0.089\%$                                                                 \\
    \bottomrule[1.5pt]
    \end{tabular}
    \caption{The RHF energy ($E_0$) of ${\rm{H_2}}$ at the bond length of 0.7408 \AA~(i.e., $1.4\;{\rm{a.u.}}$) using different basis sets. $N_{\rm{GTO}}$ is the total number of GTOs in each basis set regardless of whether they share the same parameters. The parameters of CDO3 are optimized to provide the lowest possible $E_0$. The results of the CBS limit are computed using a three-point extrapolation scheme~\cite{HALKIER1999437} based on the RHF results of cc-pVDZ, TZ, and QZ.}
    \label{tbCDOHF}
    \end{table}
\end{center}

As we see through the above examples, using Quiqbox, one can quickly modify an existing basis set, construct a customized basis set, and proceed with parameter optimization to meet specific research needs. In other words, Quiqbox aims to become a ``quick toolbox'' for basis set study in the NISQ era.

\section{Conclusion and perspective} \label{sec:conclusion}

On the one hand, the ideas about Gaussian basis set design and optimization~\cite{frost1967floating, dunning1977gaussian, huber1979geometry, hurley1988computation, kendall1992electron, kresse1996efficiency, tachikawa1999simultaneous, neese2011revisiting, hill2013gaussian, jensen2013atomic, white2019multisliced} have been constantly developing along the advance of classical ab initio computation for electronic structure problems. On the other hand, the existing popular Gaussian basis sets used in classical computations do not provide a smooth transition for NISQ devices as a practical solution to electronic structure problems. In the NISQ era, both advanced classical methods and potential quantum solutions are being presented and compared to draw the future outlines of ab initial electronic structure computation~\cite{cerezo_variational_2021, tilly2022variational, anand2022quantum}. Under such a premise, tailoring new basis sets for specific scenarios becomes more in demand. Aside from the need for a more sophisticated construction of basis sets, we explained the significance of a more systematic study of basis sets beyond providing optimal numerical results for specific many-electron systems in Sec.~\ref{sec:intro-nisq}. Customizing and modifying basis sets to a higher degree across different systems will provide finer control of the electronic structure problem as a general computational problem and help study the effect of basis set discretization of electronic Hamiltonians.

To better aid the study of the aforementioned subjects, we proposed a highly customizable basis set generation framework based on mix-contracted Gaussian-type orbitals (MCGTOs), which form a superset of contracted Gaussian-type orbitals (CGTOs). Moreover, this framework allows connecting orbital parameters through mappings from primitive parameters to capture the correlation among basis functions or impose extra symmetry on basis sets. This framework contains two main parts: basis set generation and parameter optimization. We introduced in Sec.~\ref{sec:bsDesign} the concept of MCGTO and explained its capability to go beyond AOs. In Sec.~\ref{sec:optPars}, we then introduced a variational parameter optimization procedure for the basis sets composed of MCGTOs, which can be implemented with general electronic structure algorithms. This framework is already realized by the open-source software package, Quiqbox~\cite{Quiqbox2022}, which we have developed using the programming language Julia~\cite{Bezanson_Julia_A_fresh_2017, bezanson2018julia}. To demonstrate the potential application of this framework as well as the accompanying package Quiqbox, we provided several examples. They range from studying the performance of parameter-optimized atomic basis sets for the HF methods (Sec.~\ref{sec:AbsOpt}) and VQE computation (Sec.~\ref{sec:VQE}) to generating and optimizing a basis set composed of ``completely delocalized orbitals'' (CDOs). A CDO basis set with three orbitals and four unique floating GTOs (CDO-4G) provides an improvement over 0.003 Ha for the HF energy of $\rm{H_2}$ from the aug-cc-pVDZ basis set consisting of 18 orbitals (see Table~\ref{tbCDOHF}). 

Just as the potential of MCGTO and Quiqbox as the first step to a more systematic study on basis sets for electronic structure is presented in this paper, more unanswered questions that require further investigation also lie ahead. We want to divide the future directions into two main categories, the study regarding the advantages of specific finite basis set configurations and the study of their CBS limit behaviors.

The first direction comes from the design of our modular optimization procedure (see FIG.~\ref{fig:po}), from which the optimization can be tailored for specific electronic structure properties and algorithms. By adding more post-HF methods as objective functions and more tests on systems of interest, we can deepen our understanding of the significance of differentiable basis set optimization for electronic structure case by case. Furthermore, we can study the correlation between basis set geometry and the performance of corresponding algorithms. For instance, given a ground state ansatz, can one design a basis set that provides low-error discretization of the target electronic Hamiltonian and reduces the computational cost of finding the ground state? If such a basis set exists, then the parameter space characterized by the basis set will contain a global minimum closer to the true ground-state wavefunction and allows the optimizer to find it easily. In this case, the objective of the basis set design is not necessarily tied to a specific system. Therefore, instead of optimizing the basis set parameters with respect to a specific system, we should optimize them with respect to an ensemble of molecules. A more detailed discussion relating basis set design and complexity of electronic structure is presented in Sec.~\ref{sec:intro-nisq}.

Additionally, it is important to connect the performance of a specific basis set construction to its performance approaching the CBS limit~\cite{peterson2007gaussian}. This leads to our second category of future directions. The basis set errors, such as basis set truncation error (BSTE) and basis set superposition error (BSSE) from local basis functions, always exist for a finite (incomplete) basis set regardless of the optimization of basis set parameters. There are methods to handle specific sources of error, such as the counterpoise correlation for BSSE~\cite{cook1993some}. However, to eliminate these errors altogether, one would need to achieve the CBS limit. A basis set family having a systematic convergence with respect to a certain physical observable allows extrapolation of its CBS limit performance without having to compute the observable at many different basis set sizes~\cite{HALKIER1999437}. Unfortunately, this property does not come automatically for any basis set design, which is why correlation-consistent basis sets are a popular choice for post-HF computation~\cite{truhlar1998basis, peterson2007gaussian}. With Quiqbox, one can optimize the basis set with respect to a particular system. Specifically, we considered the case of optimizing atomic basis sets in Sec.~\ref{sec:AbsOpt}. If follow-up research can determine the convergence behavior of re-optimized cc-pVxZ (where x controls the basis set size), then a system-specific CBS extrapolation can be achieved in combination with basis set parameter optimization. Furthermore, we can try incorporating new basis set completeness optimization techniques into the basis set construction and optimization procedure~\cite{manninen2006systematic, lehtola_2014}. This may allow us to provide a supplementary CBS extrapolation scheme.

Last but not least, we would like to continue optimizing Quiqbox's performance. In this way, besides being a basis set generation and optimization packages, it can also become a stronger option for doing related ab initio electronic structure computation.

In the era where both NISQ and classical methods for electronic structure are advancing, we proposed a framework with higher customizability and flexibility to approach basis set design. We hope this framework, with the software Quiqbox we have developed, is useful in extending existing methods in quantum chemistry and beyond.

% \section{Acknowledgements}
\begin{acknowledgement}
This paper was supported by the ``Quantum Chemistry for Quantum Computers'' project sponsored by the U.S. Department of Energy under Award No. DESC0019374. The authors are thankful for Casey Dowdle's assistance with the VQE computation mentioned in Sec.~\ref{sec:VQE} and Brent Harrison's proofreading and comments on the initial draft. We are grateful for the anonymous referees' constructive comments and questions, which improved the completeness of our research and the readability of this paper. JDW was supported by the US NSF grant PHYS-1820747 and NSF EPSCoR-1921199 and by the Office of Science, Office of Advanced Scientific Computing Research under programs Fundamental Algorithmic Research for Quantum Computing (FAR-QC) and the Optimization, Verification, and Engineered Reliability of Quantum Computers (OVER-QC) projects. JDW holds concurrent appointments at Dartmouth College and as an Amazon Visiting Academic. This paper describes work performed at Dartmouth College and is not associated with Amazon. WW would like to express his gratitude to his mother, father, and sister. This paper would not have been possible without their utmost support. He would also like to thank his friends for their constant encouragement.
\end{acknowledgement}

\bibliography{main}

\newpage

\begin{appendices}

\section{Quiqbox performance} \label{app:performance}

To demonstrate the performance of Quiqbox (version 0.5.7 on Julia 1.8.5), we compared it in three scenarios with existing packages that share similar features to a limited extent. The tests were run on a cluster node equipped with an AMD EPYC 7532 CPU.

\subsection{The Hartree--Fock method} \label{app:bm-HF}

As the HF method is currently the main objective function for basis set optimization in Quiqbox, it is crucial to verify the accuracy and efficiency of Quiqbox in performing such computations. We selected various types of molecules based on their geometries and chemical bonds, including hydrogen chains, homonuclear diatomic molecules, hydrides, and metallic compounds. The distance between adjacent nuclei of hydrogen chains is set to $1.0$ a.u., and the geometries of other molecules are set to the experimental geometries recorded on CCCBDB~\cite{cccbdb}. 

As for the reference software package, we chose the widely-used ab initio electronic structure package PySCF~\cite{sun2020recent} (version 2.2.1 running on Python 3.10.11).  The convergence methods used in the two packages remained in their default configuration. The convergence thresholds of the SCF iteration on two packages were both set to $1\!\times\!10^{-6}\,\rm{Ha}$ and $1\!\times\!10^{-4}\,\rm{a.u.}$ for the step-wise changes of energy and the orbital rotation gradient~\cite{pulay1982improved, hu2010accelerating, helgaker2013molecular}, respectively. The initial guess of the charge density matrix in both packages was constructed from the orbital coefficient matrix obtained by diagonalizing the core Hamiltonian of the system (i.e., the ``core guess''). In this way, we could have a more aggressive test on the default convergence procedures in both packages. 

Additionally, there were a few manual modifications we needed to make in order to provide a more proper comparison of the HF SCF computation between the two packages. First, we found that, on the one hand, the basis functions stored in PySCF have optimized general contraction coefficients, which in some cases have fewer GTOs than the standard versions stored in Quiqbox for the same basis set. On the other hand, the version of these basis sets was older than those stored in Quiqbox, such that the regression they caused in the HF energy computation was non-negligible. So, we replaced the version of basis sets in Quiqbox with the one used in PySCF for benchmarking. Secondly, the option \texttt{direct\_scf} in PySCF was disabled as it enforces the construction of the Fock matrix to be an extrapolation of the Fock matrices from previous iterations rather than a rigorous computation. Finally, the multi-threading computation was enabled for both packages, with the total allowed number of threads set to eight. A code snippet of the two packages' configuration for the HF benchmark is shown in FIG.~\ref{fig:HFcode}.

\begin{figure}[htbp]
    \centering
    \begin{subfigure}[b]{0.48\textwidth}
        \centering
        {\normalsize \jlinputlisting[]{QuiqboxRHF.jl}}
        \caption{The Julia code of Quiqbox's implementation (in version 0.5.7). It is worth noting that the user would typically use function \texttt{genBasisFunc} to construct basis functions either from GTOs or atomic basis set names (like specifying \texttt{bsKey} in FIG.~\ref{lst:pHF}). However, due to the inconsistency of the default atomic basis versions stored in two packages, we applied \texttt{genBFuncsFromText} to generate the exact same basis sets as the ones used in PySCF.}
        \label{lst:qHF}
     \end{subfigure}
     \hfill
     \begin{subfigure}[b]{0.48\textwidth}
        \centering
        {\normalsize \jlinputlisting[]{PySCFRHF.py}}         
        \caption{The Python code of PySCF's implementation (in version 2.2.1). A helper function \texttt{save\_scf\_iteration} is implemented as this version of PySCF does not natively support exporting the cycle information of SCF iterations.}
        \label{lst:pHF}
     \end{subfigure}
     \caption{The code for implementing RHF from Quiqbox and PySCF, respectively. The additional wrapper functions and code for the benchmarking routine are omitted. For more information on the functions used in two pieces of code, we refer the reader to the official documentation of the two packages~\cite{Quiqbox2022, sun2020recent}.}
    \label{fig:HFcode}
\end{figure}

\begin{figure*}[htbp]
    \centering
    \begin{subfigure}[b]{0.4\textwidth}
         \centering
         \includegraphics[width=0.99\textwidth]{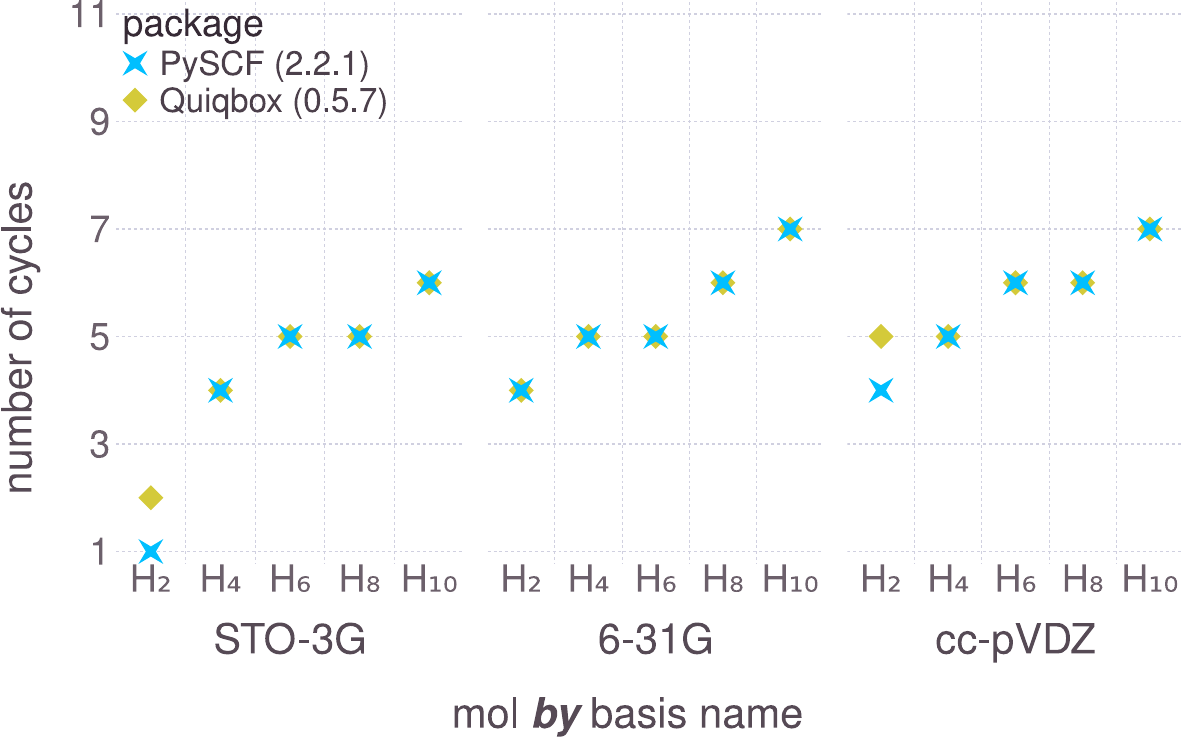}
         \caption{}
         \label{fig:bm_HF_nI_HC}
     \end{subfigure}
     \hspace{0.5em}
     \begin{subfigure}[b]{0.4\textwidth}
         \centering
         \includegraphics[width=0.99\textwidth]{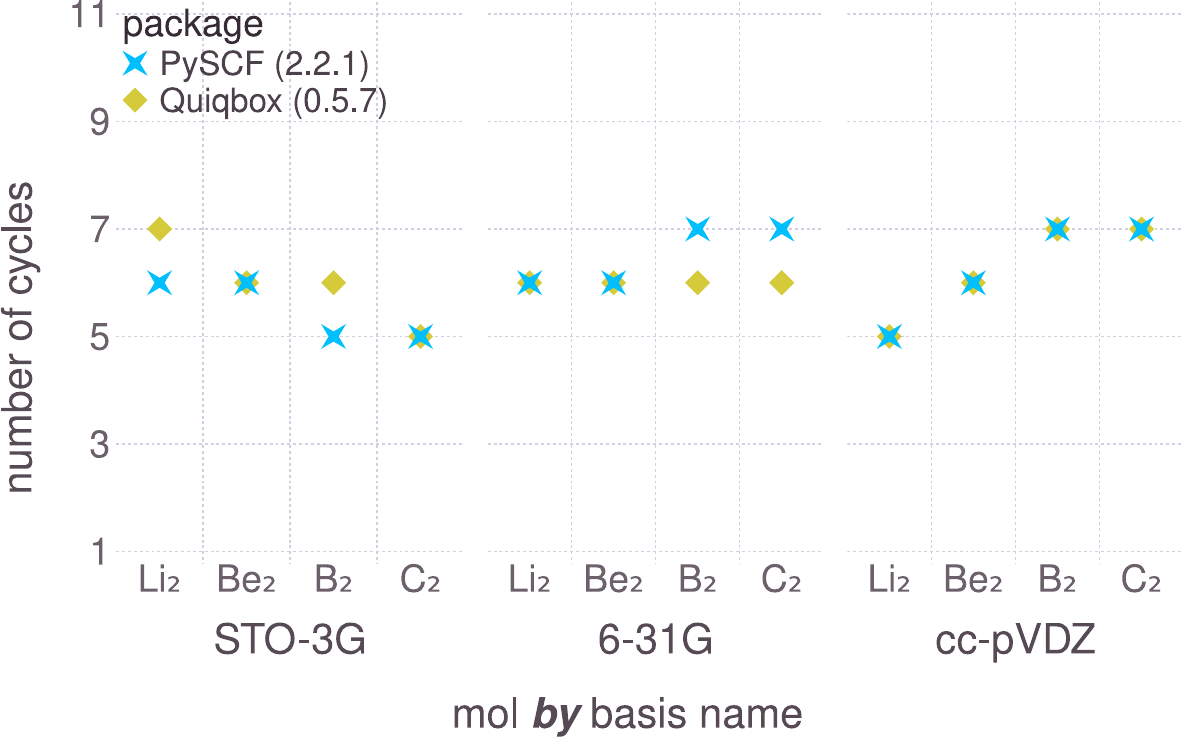}
         \caption{}
         \label{fig:bm_HF_nI_DHM}
     \end{subfigure}
     \begin{subfigure}[b]{0.4\textwidth}
         \centering
         \includegraphics[width=0.99\textwidth]{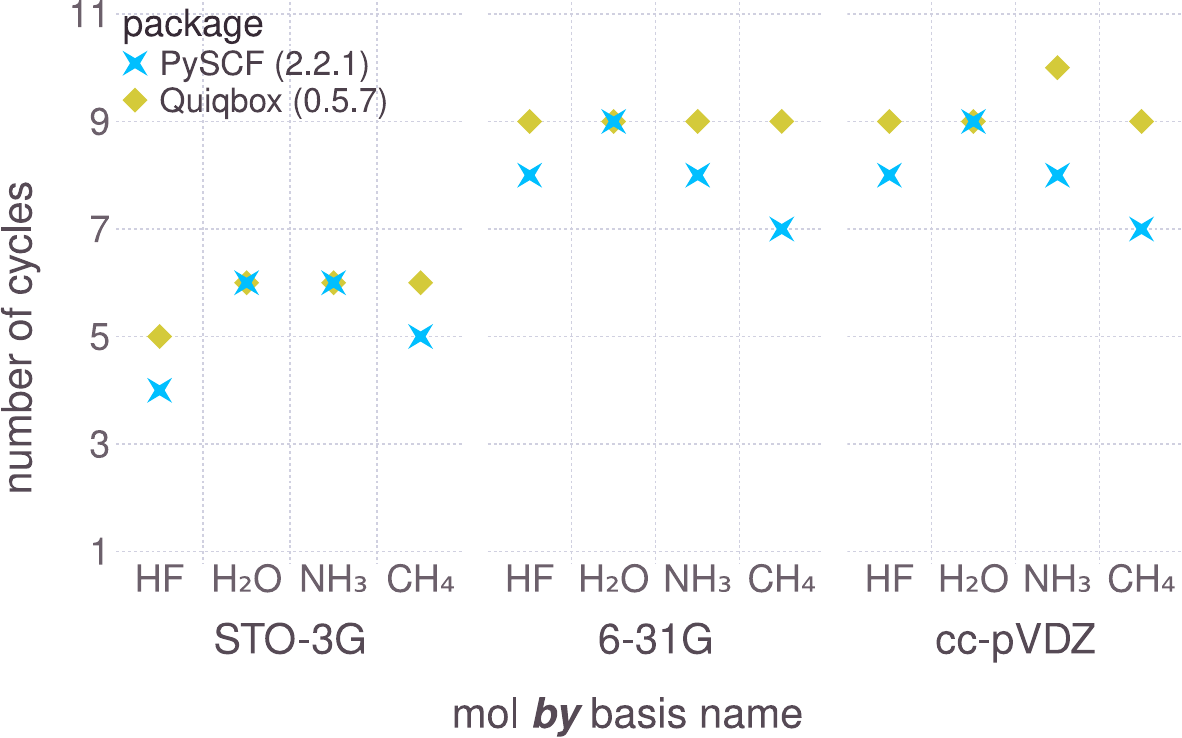}
         \caption{}
         \label{fig:bm_HF_nI_XH}
     \end{subfigure}
     \hspace{0.5em}
     \begin{subfigure}[b]{0.4\textwidth}
         \centering
         \includegraphics[width=0.99\textwidth]{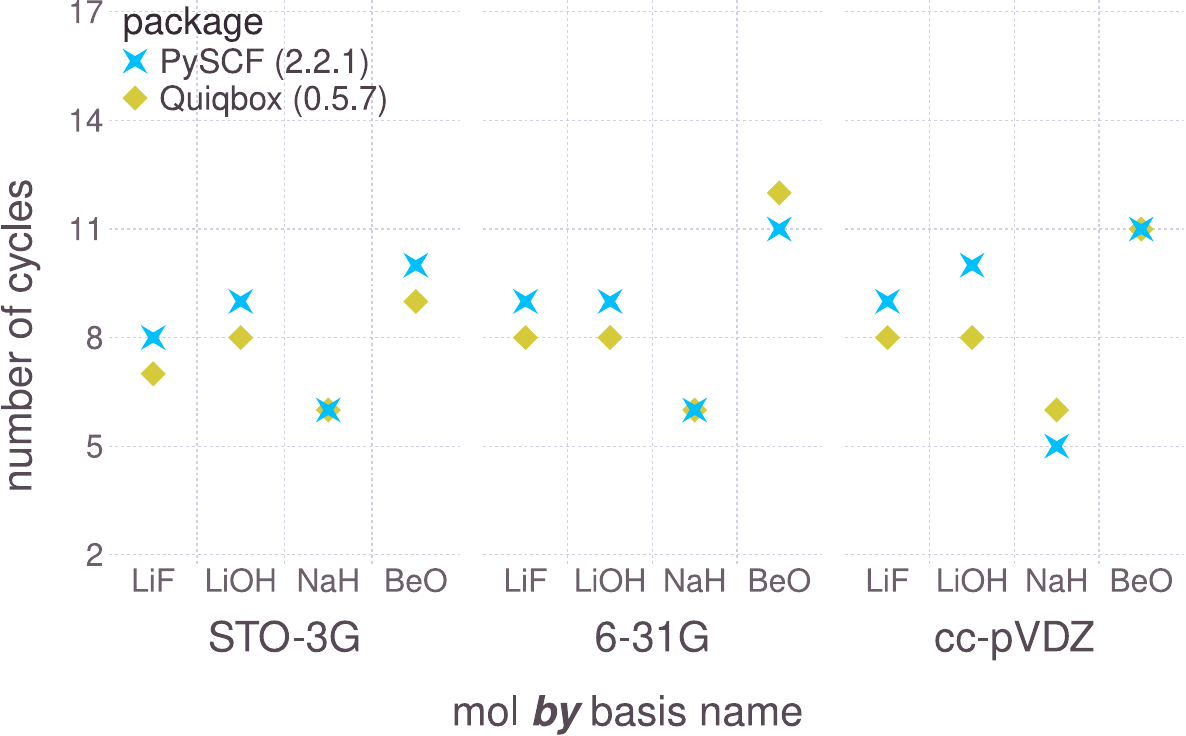}
         \caption{}
         \label{fig:bm_HF_nI_MC}
     \end{subfigure}
    \caption{The number of cycles needed to converge the RHF energy of various molecular systems using SCF procedure with different basis sets.}
    \label{fig:bm_HF_nI}
\end{figure*}

\begin{figure*}[htbp]
    \centering
    \begin{subfigure}[b]{0.4\textwidth}
         \centering
         \includegraphics[width=0.99\textwidth]{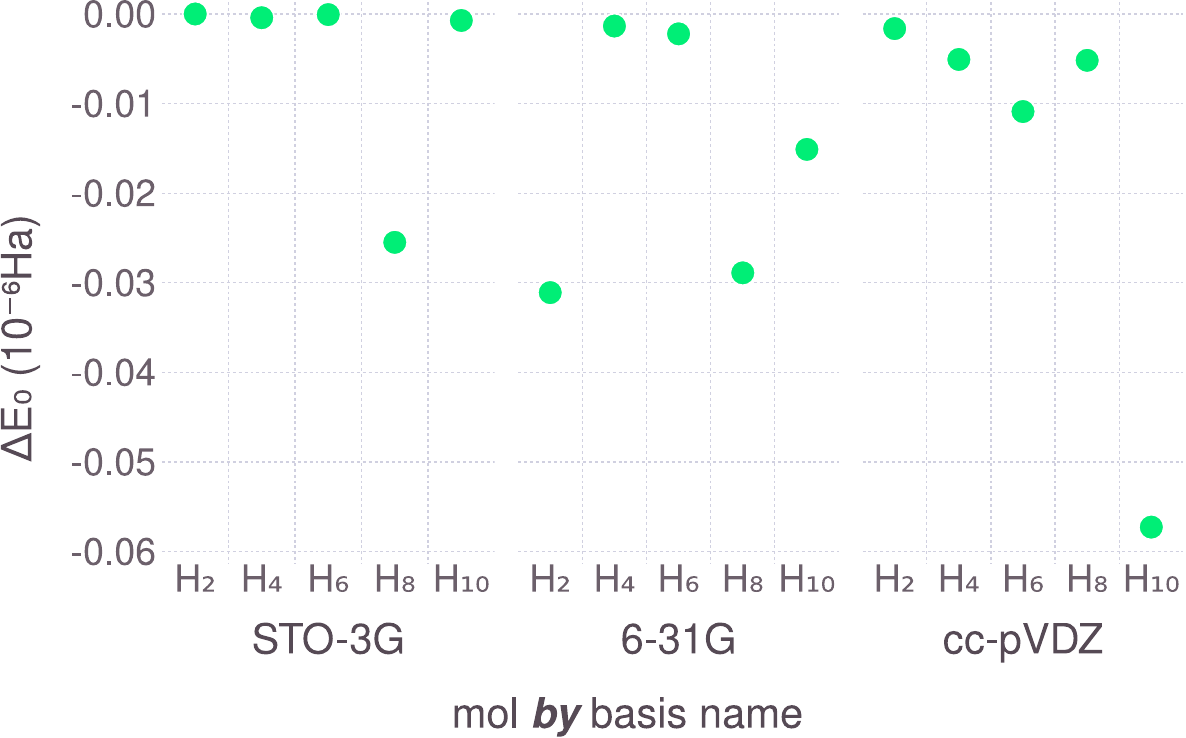}
         \caption{}
         \label{fig:bm_HF_dE_HC}
     \end{subfigure}
     \hspace{0.5em}
     \begin{subfigure}[b]{0.4\textwidth}
         \centering
         \includegraphics[width=0.99\textwidth]{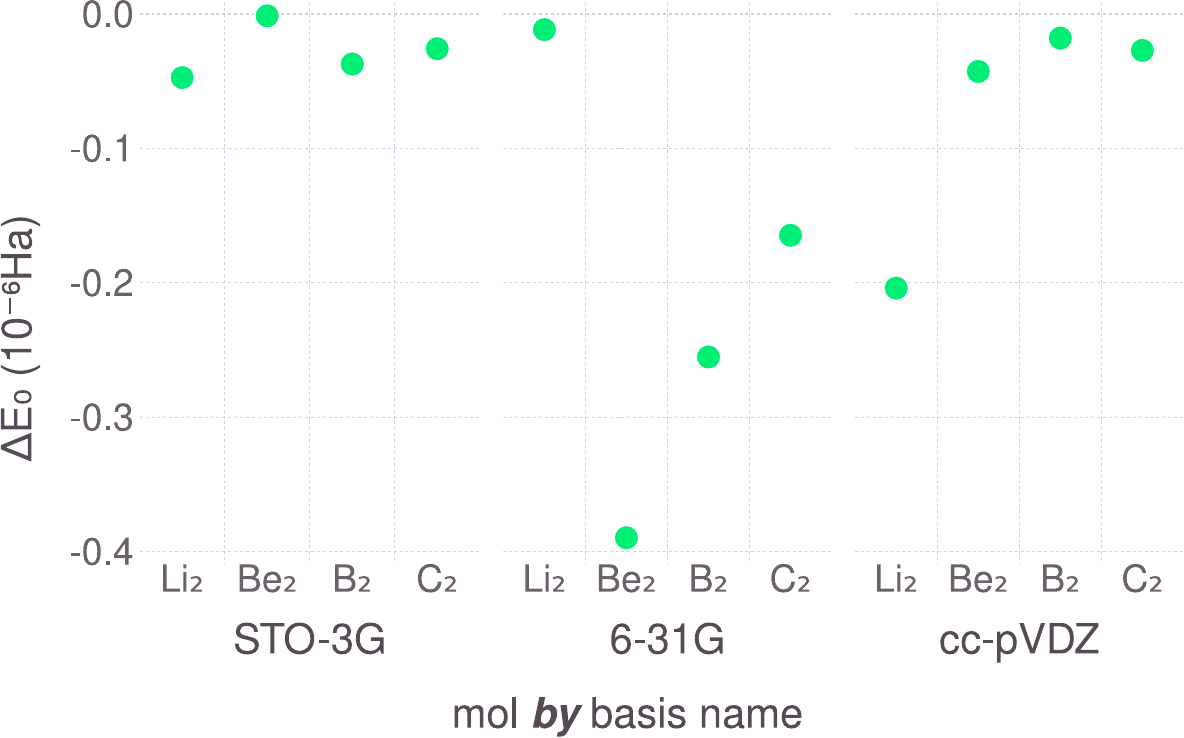}
         \caption{}
         \label{fig:bm_HF_dE_DHM}
     \end{subfigure}
     \begin{subfigure}[b]{0.4\textwidth}
         \centering
         \includegraphics[width=0.99\textwidth]{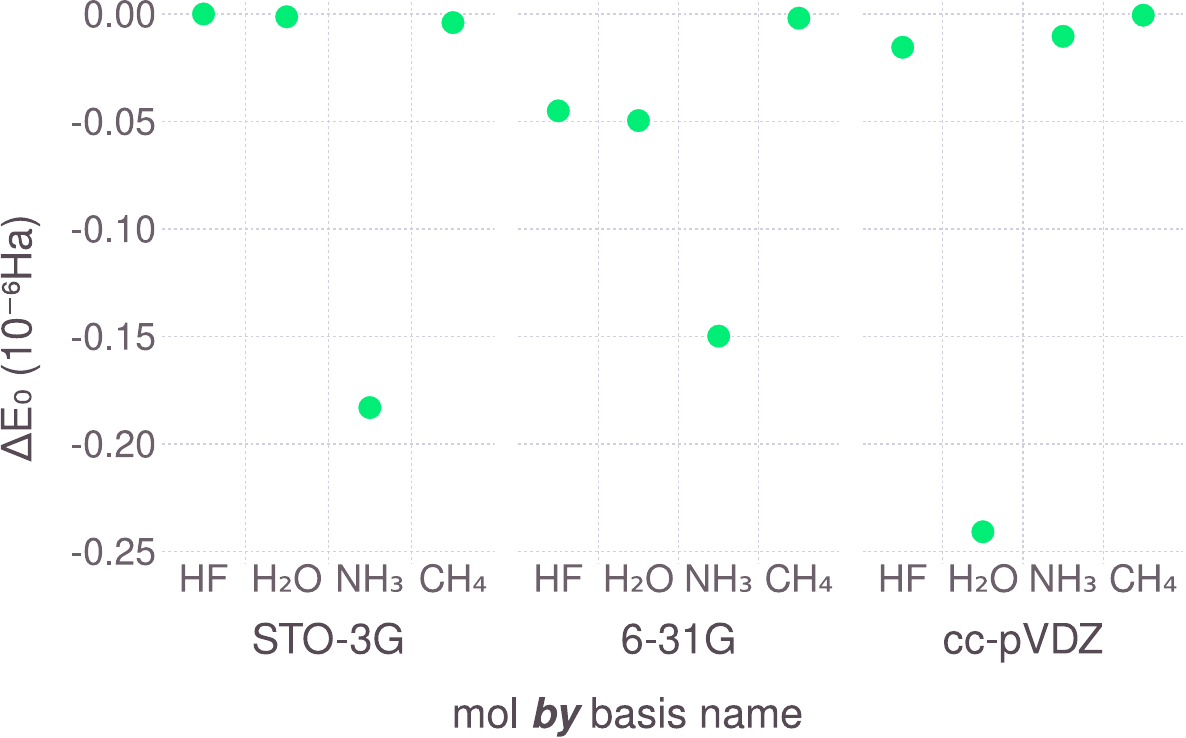}
         \caption{}
         \label{fig:bm_HF_dE_XH}
     \end{subfigure}
     \hspace{0.5em}
     \begin{subfigure}[b]{0.4\textwidth}
         \centering
         \includegraphics[width=0.99\textwidth]{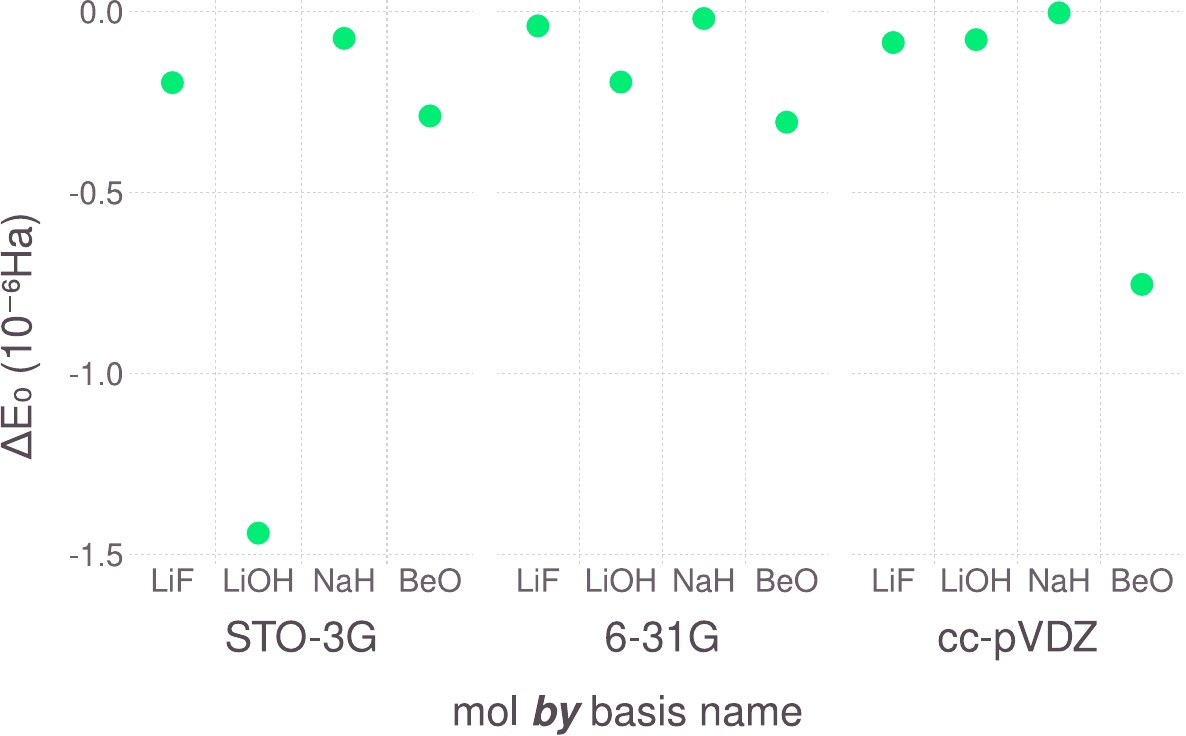}
         \caption{}
         \label{fig:bm_HF_dE_MC}
     \end{subfigure}
    \caption{The converged RHF energy of Quiqbox's SCF iteration compared to PySCF's for various molecular systems with different basis sets. $\Delta E_0$ represents the converged RHF energy from PySCF minus the one from Quiqbox.}
    \label{fig:bm_HF_dE}
\end{figure*}

\begin{figure*}[htbp]
    \centering
    \begin{subfigure}[b]{0.4\textwidth}
         \centering
         \includegraphics[width=0.99\textwidth]{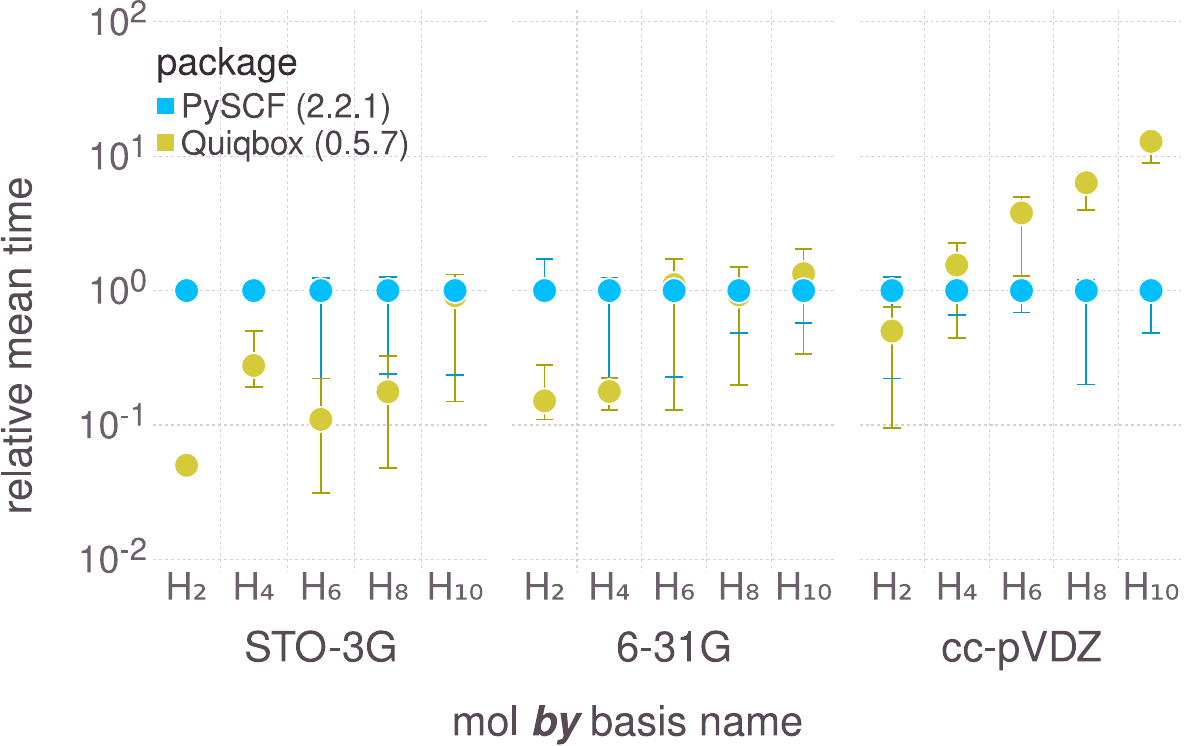}
         \caption{}
         \label{fig:bm_HF_t_HC}
     \end{subfigure}
     \hspace{0.5em}
     \begin{subfigure}[b]{0.4\textwidth}
         \centering
         \includegraphics[width=0.99\textwidth]{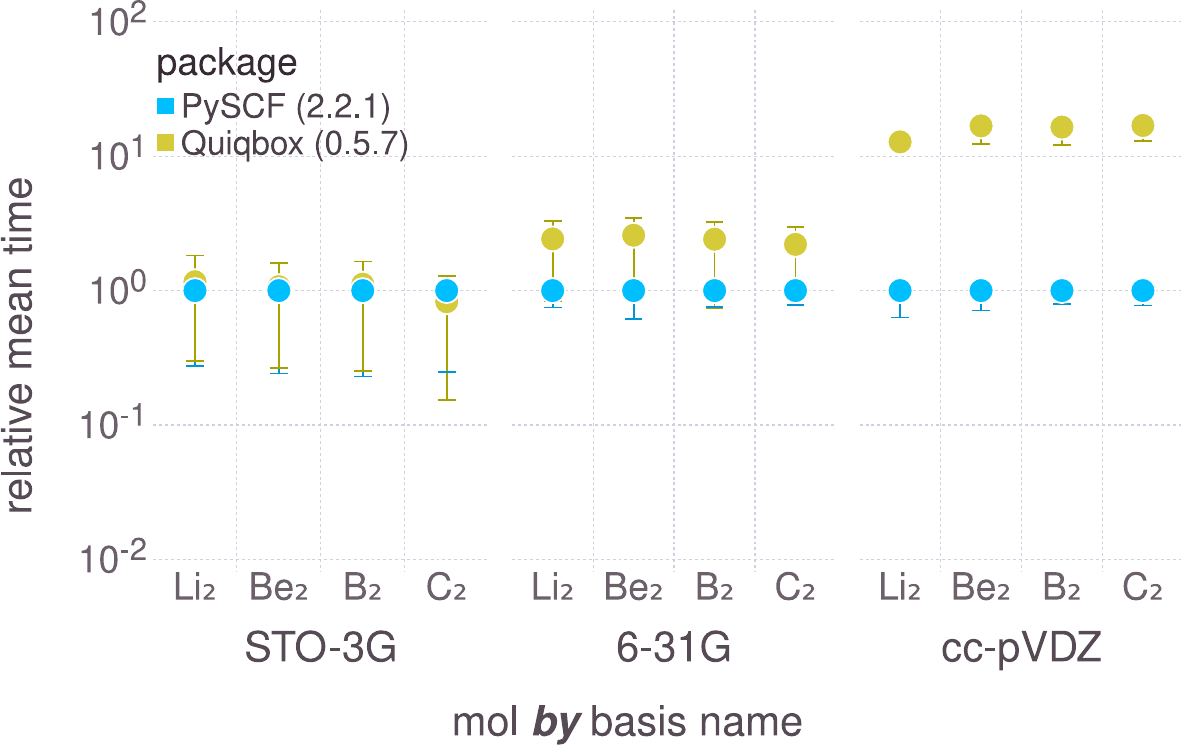}
         \caption{}
         \label{fig:bm_HF_t_DHM}
     \end{subfigure}
     \begin{subfigure}[b]{0.4\textwidth}
         \centering
         \includegraphics[width=0.99\textwidth]{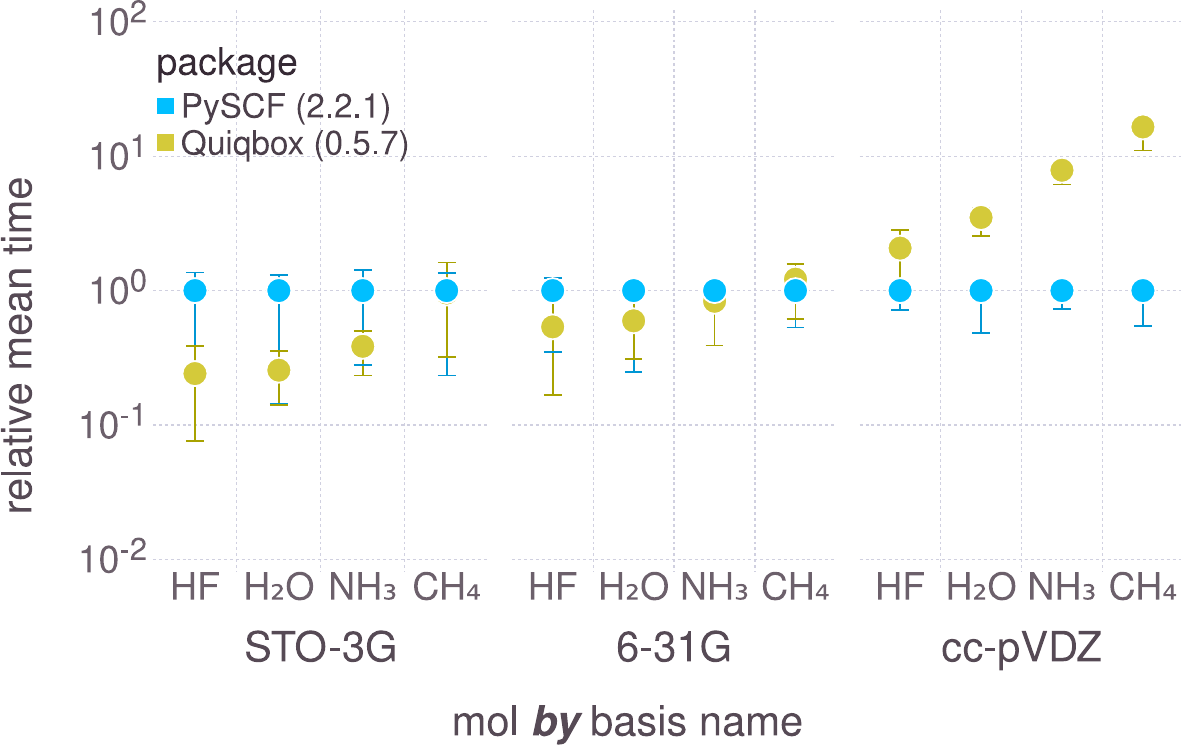}
         \caption{}
         \label{fig:bm_HF_t_XH}
     \end{subfigure}
     \hspace{0.5em}
     \begin{subfigure}[b]{0.4\textwidth}
         \centering
         \includegraphics[width=0.99\textwidth]{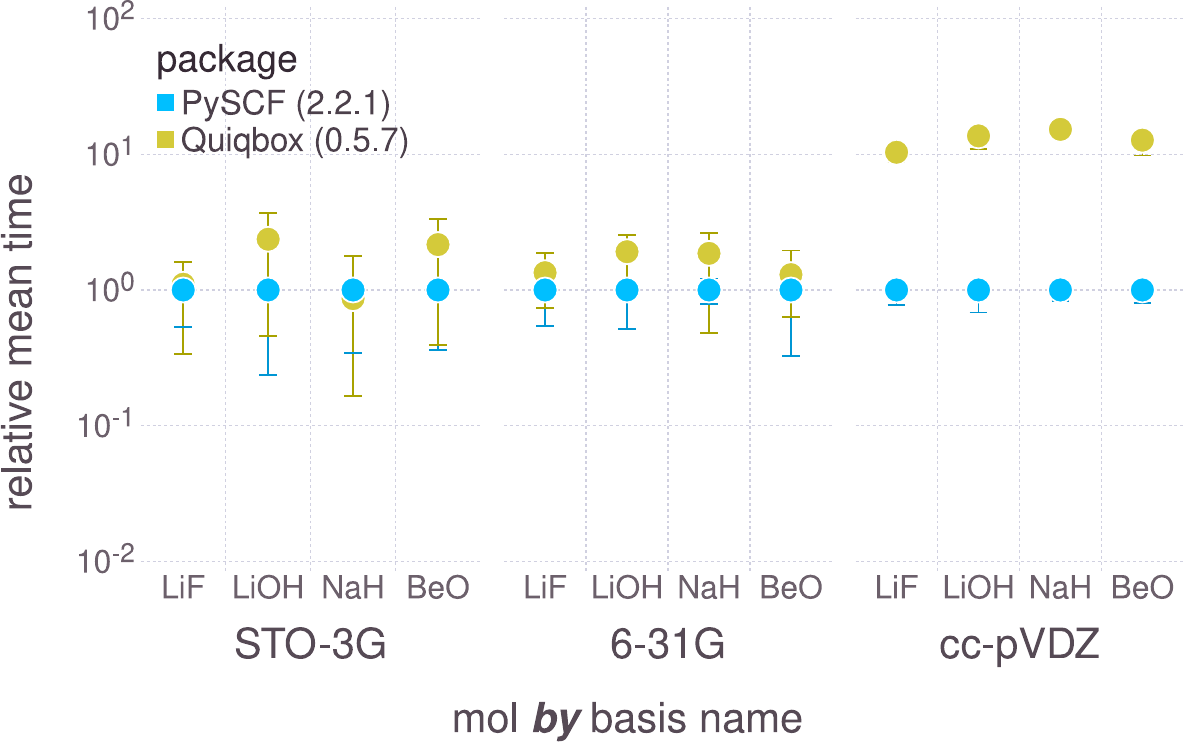}
         \caption{}
         \label{fig:bm_HF_t_MC}
     \end{subfigure}
    \caption{The relative mean runtimes of Quiqbox's RHF SCF iteration compared to PySCF's. The upper error bar indicates the mean time plus one standard deviation, and the lower error bar records the minimal runtime.}
    \label{fig:bm_HF_t}
\end{figure*}

The benchmarking results are shown in FIG.~\ref{fig:bm_HF_nI}--\ref{fig:bm_HF_t}. Specifically, FIG.~\ref{fig:bm_HF_nI} shows the number of cycles needed for the SCF iteration to converge, where we can see that Quiqbox and PySCF are roughly in agreement as the numbers of cycles for two packages do not differ from each other by more than two. The computed RHF energies from the two packages are also consistent with each other, as shown in FIG.~\ref{fig:bm_HF_dE}. Most of the differences between the converged energy are below the convergence threshold, with the only exception being the case of LiOH (STO-3G) in FIG.~\ref{fig:bm_HF_dE_MC}, where the difference was around $1.4\!\times\!10^{-6}$ Ha that is still below the chemical accuracy.

In FIG.~\ref{fig:bm_HF_t}, we show that the ratios of the average runtimes of Quiqbox's and PySCF's RHF procedure are mostly within the same magnitude for basis sets STO-3G and 6-31G. Sometimes, e.g., for hydrogen chains and hydrides, Quiqbox outperformed PySCF. However, when the larger basis set cc-pVDZ was applied, the ratio of Quiqbox's runtime to Libcint was increased to, at most, around 10. The reason for this is that the recorded RHF runtimes include the time to compute electronic integral for both packages and currently, Quiqbox's own integral engine scales worse than PySCF's.

\subsection{Electronic integrals} \label{sec:Int}

Instead of directly relying on an external Gaussian-type electronic integral library, we wrote our own integral engine (also in Julia) as part of the core functions of Quiqbox, based on references~\cite{petersson2009detailed, fermann2020fundamentals}. Additionally, we have added specialized optimization methods to avoid repetitive integral computation resulting from the orbitals with shared parameters. On top of that, the integral engine is fully compatible with MCGTOs. By contrast, PySCF relies on the C-based electronic integral library Libcint~\cite{libcint2015}. 

We compared the basic electronic integral computations of the overlap matrix, the core Hamiltonian, and the electron--electron interaction between Quiqbox and Libcint. We used the hydrogen chains (with up to 18 hydrogen atoms) again to test how the computation accuracy and efficiency scale with respect to the growth of system size. Since Quiqbox does not have the feature of automatically optimizing the computation based on the symmetry of the molecule as PySCF, we manually turned off the option in PySCF. Again, due to the issue with PySCF's pre-stored basis sets mentioned in Appx.~\ref{app:bm-HF}, we performed the same replacement for Quiqbox's basis sets. 

\begin{figure*}[htbp]
    \centering
    \begin{subfigure}[b]{0.4\textwidth}
        \centering
        \includegraphics[width=0.99\textwidth]{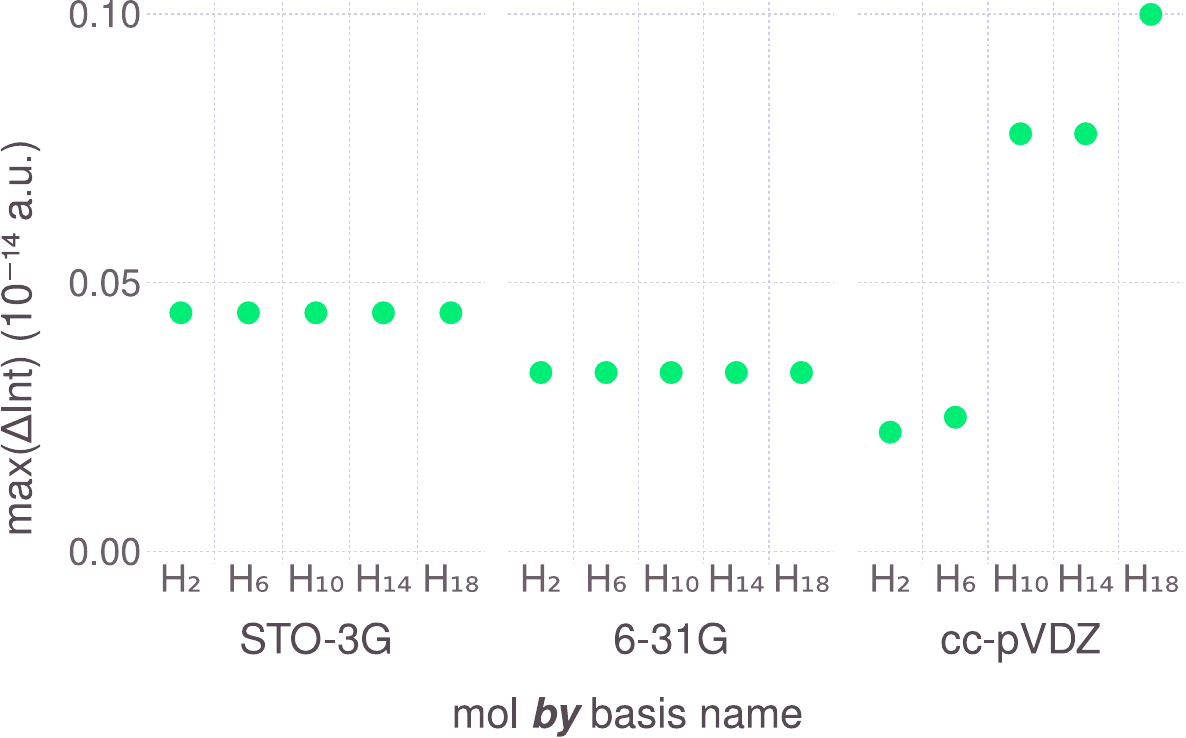}
    \caption{Overlap matrix.}
        \label{fig:bm_int_err_overlap}
    \end{subfigure}
    \begin{subfigure}[b]{0.4\textwidth}
        \centering
        \includegraphics[width=0.99\textwidth]{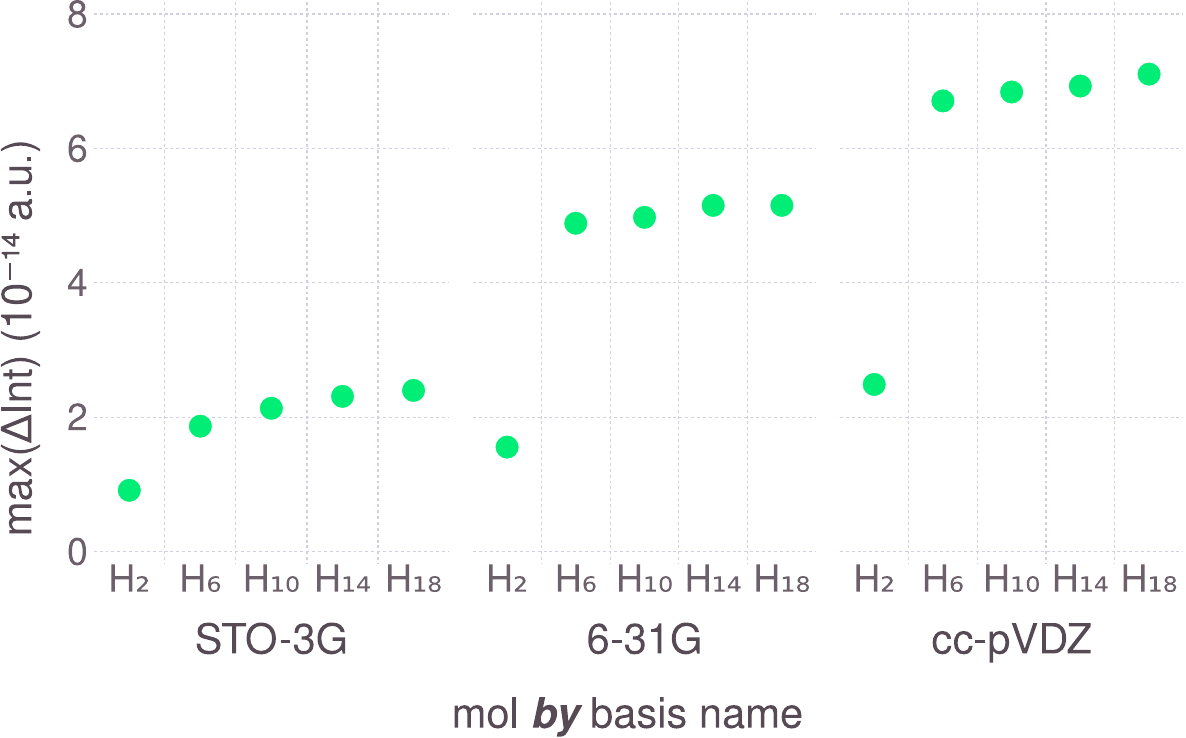}
    \caption{Core Hamiltonian.}
        \label{fig:bm_int_err_coreH}
    \end{subfigure}
    \begin{subfigure}[b]{0.4\textwidth}
        \centering
        \includegraphics[width=0.99\textwidth]{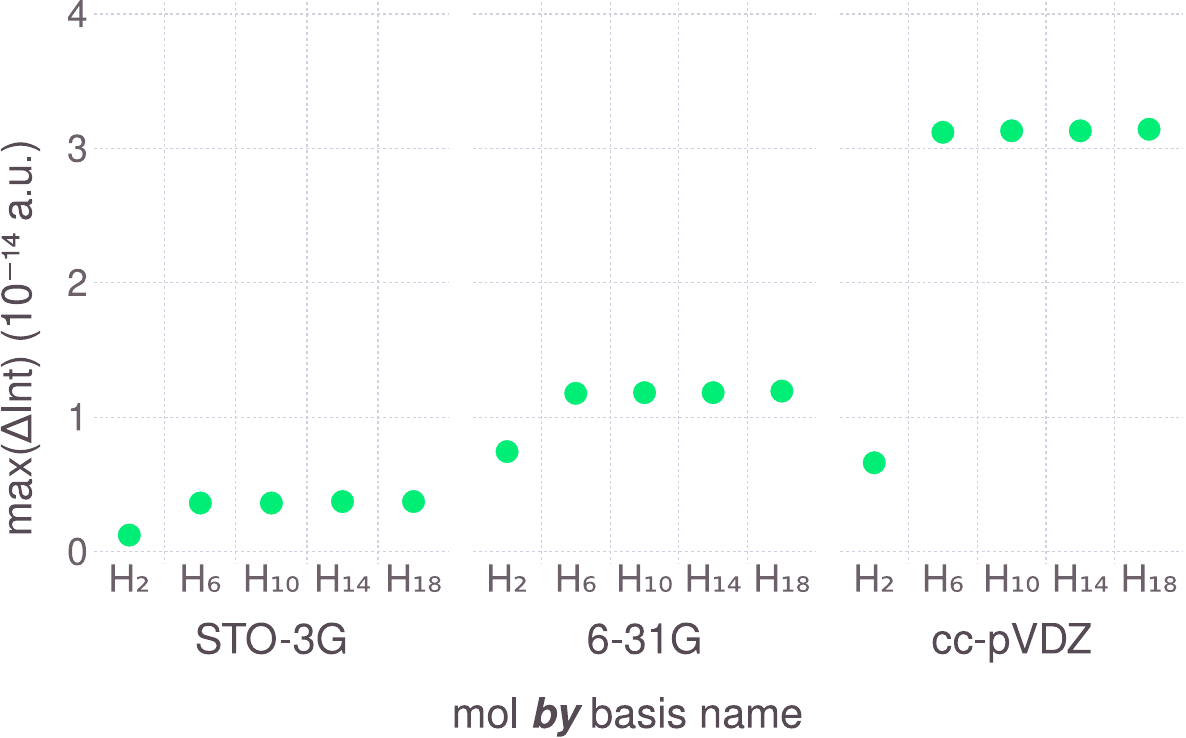}
        \caption{electron--electron interaction.}
        \label{fig:bm_int_err_eeI}
    \end{subfigure}
    \caption{The element-wise differences between the same electronic integral tensors computed by either Quiqbox or Libcint. max($\Delta$Int) represents the maximal value of the element-wise differences for a specific hydrogen chain.}
    \label{fig:bm_int_err}
\end{figure*}

\begin{figure*}[htbp]
    \centering
    \begin{subfigure}[b]{0.4\textwidth}
        \centering
        \includegraphics[width=0.99\textwidth]{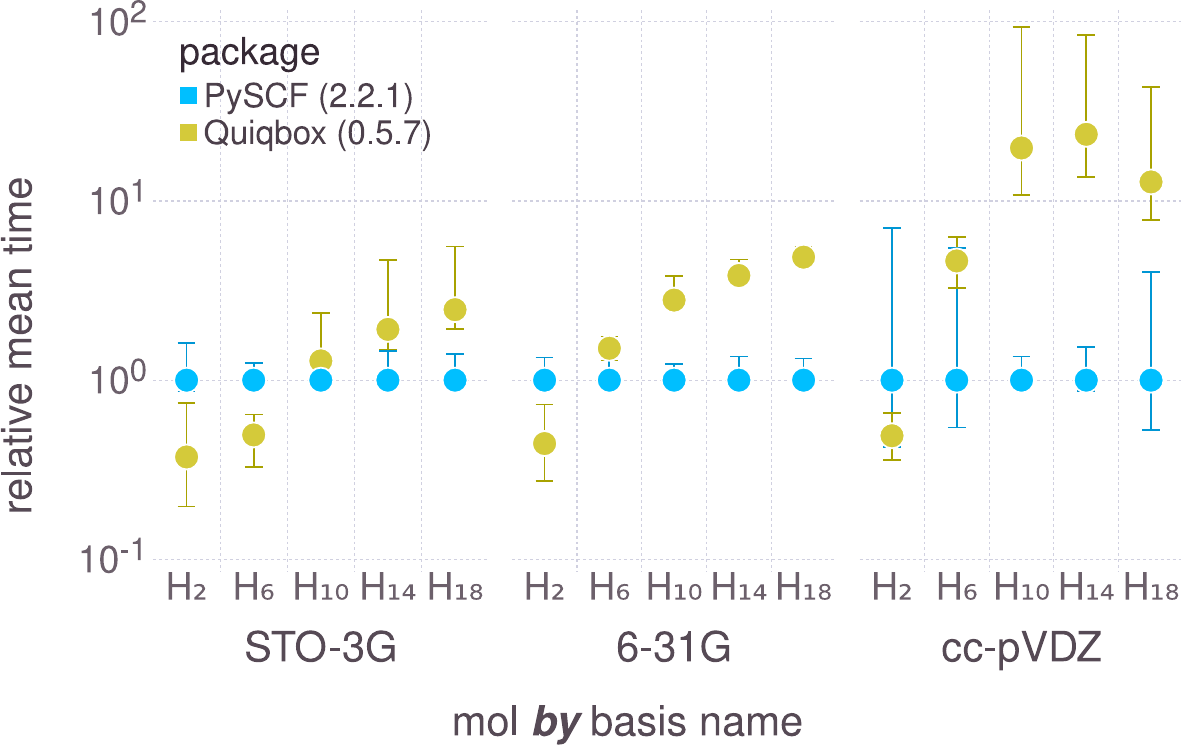}
    \caption{Overlap matrix.}
        \label{fig:bm_int_t_overlap}
    \end{subfigure}
    \begin{subfigure}[b]{0.4\textwidth}
        \centering
        \includegraphics[width=0.99\textwidth]{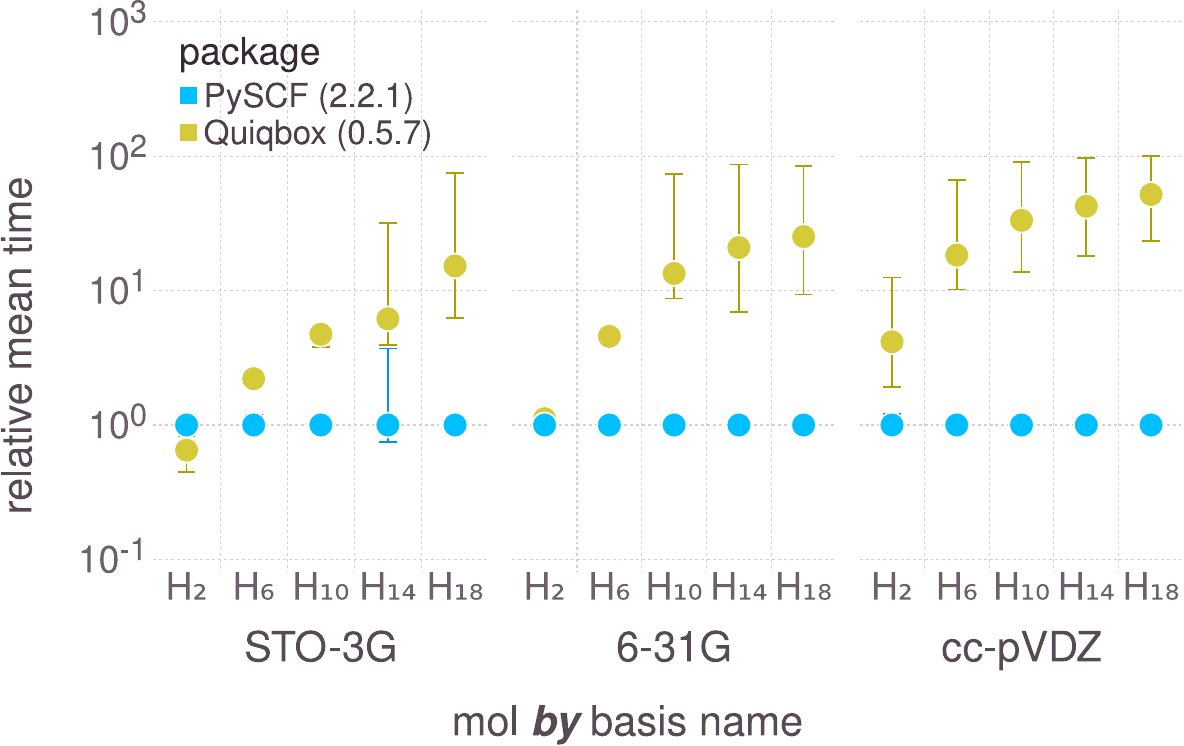}
    \caption{Core Hamiltonian.}
        \label{fig:bm_int_t_coreH}
    \end{subfigure}
    \begin{subfigure}[b]{0.4\textwidth}
        \centering
        \includegraphics[width=0.99\textwidth]{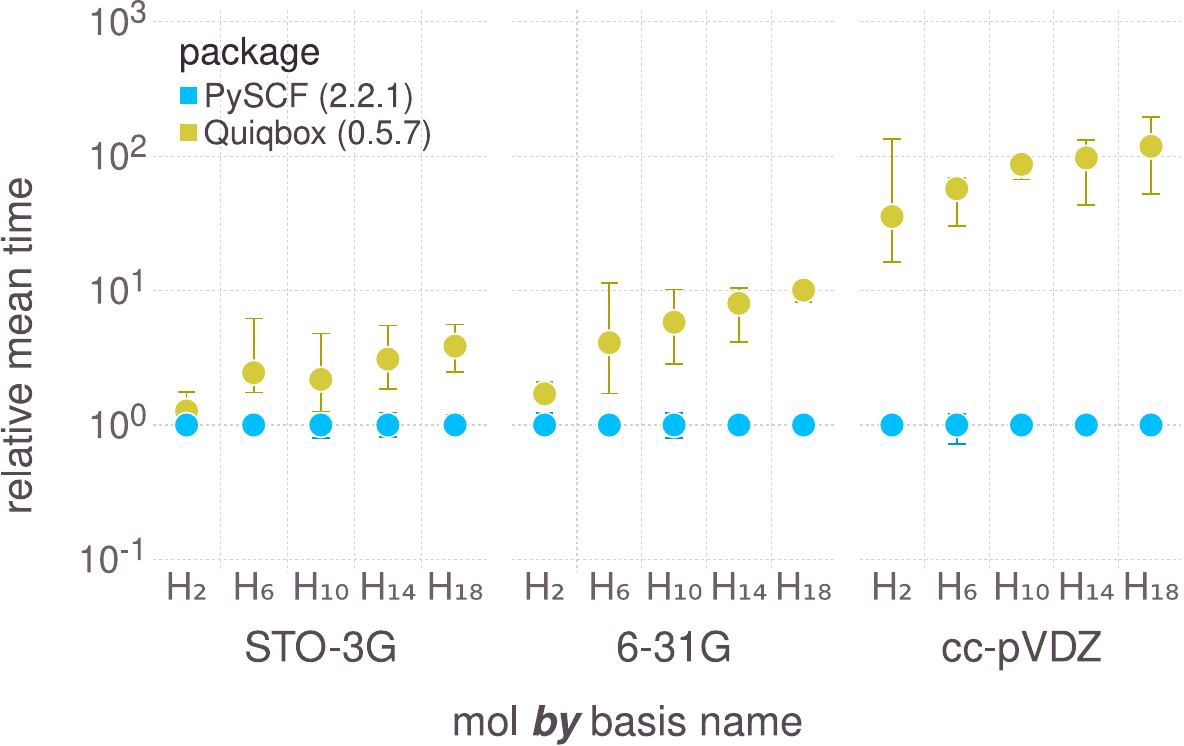}
        \caption{electron--electron interaction.}
        \label{fig:bm_int_t_eeI}
    \end{subfigure}
    \caption{Quiqbox's relative mean runtimes of computing electronic integrals for different hydrogen chains with respect to Libcint's. Same as in FIG.~\ref{fig:bm_HF_t}, the upper error bar indicates the mean time plus one standard deviation, and the lower error bar records the minimal runtime.}
    \label{fig:bm_int_t}
\end{figure*}

Quiqbox and Libcint agreed on all the computation results within a difference of $10^{-13}$ a.u., as shown in FIG.~\ref{fig:bm_int_err}. In respect of computation efficiency, we can see from FIG.~\ref{fig:bm_int_t}, for smaller basis sets such as STO-3G and 6-31G, the minimal runtime difference between Quiqbox's integral engine and Libcint is within the same magnitude in best cases. For the cc-pVDZ basis set, the differences increase, particularly for electron--electron interaction integrals. The author of Libcint mentioned in their technical literature that they implemented optimization on the hardware level for processes such as CPU caching and addressing, as well as software-level optimization for sparse matrix computation to improve Libcint's overall performance~\cite{libcint2015}, both of which are currently lacking in Quiqbox's integral engine. We believe that by implementing this optimization in the future, we can shorten the efficiency gap between Quiqbox's integral engine and Libcint.

Fundamentally, the benchmarks between Quiqbox and PySCF serve as a brief demonstration of Quiqbox's general user experience in terms of performance, not a strict competition against PySCF. This is not only because benchmarking software packages written in different programming languages is a complex subject but also because Quiqbox and PySCF are only partially functionally overlapped. The former is a basis set optimization package (as of the version tested), and the latter is a comprehensive quantum chemistry package. The different goals of the two packages affect their code design even when implementing the features seemingly the same on the surface level.

For example, how Quiqbox stores the basis set information differs drastically from PySCF. In the PySCF (libcint), where only pre-stored CGTO is allowed, the basis set data can be easily (and is) reduced to a homogeneous floating-point number array. On the other hand, due to the flexibility and customizability of MCGTO in Quiqbox, most tunable basis set parameters (in the custom datatype \texttt{ParamBox}) have their own meta-information and data structure. On top of this, basis functions are constructed hierarchically (primitive parameters $\to$ GTO $\to$ CGTO $\to$ MCGTO) and mutable by reassignments, combinations, separation, etc. From a high-performance computing standpoint, this non-homogeneous storing can affect the program's performance, such as integral computation. It is also worth mentioning that implementing our native integral engine allows us to realize proper normalization for Cartesian-coordinate-based GTO integrals, unlike Libcint~\cite{libcintNorm}. Maintaining the proper normalization is crucial for the MCGTOs used in Quiqbox.

\subsection{Basis set parameter optimizations} \label{sec:po}

To showcase the performance of Quiqbox's basis set optimization, we compared it with DiffiQult, one of the first open-source software packages that implemented differentiable HF methods~\cite{tamayo2018automatic}. Since DiffiQult does not support basis sets beyond atomic basis sets, the testing basis set was set to STO-3G. Moreover, DiffiQult does not support multi-threading computation, so Quiqbox was constrained to only use one thread both when running native Julia functions and when calling the default back-end linear algebra library OpenBLAS~\cite{wang2013augem}.

\begin{table*}[htbp]
    \centering
    \setlength{\tabcolsep}{2.2pt}
    \renewcommand{\arraystretch}{1.2}
    \begin{tabular}{cccccrrcrr}
        \toprule[1.5pt]
        \multirow{2}{*}{Molecule} & \multicolumn{2}{c}{Parameters} & \multirow{2}{*}{Package} & \multirow{2}{*}{Optimizer} & \multicolumn{1}{c}{\multirow{2}{*}{$E_0$ (Ha)}} & \multicolumn{1}{c}{\multirow{2}{*}{Step}} & \multicolumn{1}{c}{\multirow{2}{*}{\begin{tabular}[c]{@{}c@{}}Wall-clock\\ time (sec)\end{tabular}}} & \multicolumn{2}{c}{Speed-up} \\
         & \multicolumn{1}{l}{Types} & \multicolumn{1}{l}{Quantity} &  &  & \multicolumn{1}{c}{} & \multicolumn{1}{c}{} & \multicolumn{1}{c}{} & \multicolumn{1}{c}{Total} & \multicolumn{1}{c}{Per-step} \\ \midrule[1.2pt]\midrule[1.2pt]
        \multirow{10}{*}{$\rm H_2$} & \multirow{5}{*}{$\alpha$, $d$} & 12 & DiffiQult & BFGS (SW) & -1.83731 & 30 & 843.5 & 1.0 & 1.0 \\ \cline{3-10}
         &  & \multirow{4}{*}{12 (6)} & \multirow{4}{*}{Quiqbox} & BFGS (SW) & -1.83731 & 25 & 1.2 & 702.9 & 585.8 \\
         &  &  &  & BFGS (HZ) & -1.83731 & 17 & 1.7 & 496.2 & 281.2 \\
         &  &  &  & L-BFGS (HZ) & -1.83731 & 17 & 1.7 & 496.2 & 281.2 \\
         &  &  &  & Adam & -1.83730 & 352 & 8.0 & 105.4 & 1237.1 \\ \cmidrule[1.2pt]{2-10}
         & \multirow{5}{*}{$\alpha$, $d$, $R$} & 18 & DiffiQult & BFGS (SW) & -1.84063 & 37 & 1081.7 & 1.0 & 1.0 \\ \cline{3-10}
         &  & \multirow{4}{*}{18 (9)} & \multirow{4}{*}{Quiqbox} & BFGS (SW) & -1.84082 & 27 & 1.8 & 600.9 & 438.5 \\
         &  &  &  & BFGS (HZ) & -1.84082 & 21 & 2.8 & 386.3 & 219.3 \\
         &  &  &  & L-BFGS (HZ) & -1.84082 & 21 & 2.8 & 386.3 & 219.3 \\
         &  &  &  & Adam & -1.84081 & 375 & 10.4 & 104.0 & 1054.2 \\ \midrule[1.2pt]
        \multirow{2}{*}{LiH} & \multirow{2}{*}{$\alpha$, $d$} & \multirow{2}{*}{24} & DiffiQult & BFGS (SW) & N/A & N/A & \textgreater{}$12.00\times3600$ & 1.0 & N/A \\ \cline{4-10}
         &  &  & Quiqbox & L-BFGS (HZ) & -8.96458 & 267 & \phantom{\textgreater{}}$\phantom{1}0.82\times3600$ & \textgreater{}$14.6$ & N/A \\
         \bottomrule[1.5pt]
    \end{tabular}
    \caption{The benchmark for Quiqbox's parameter optimization of STO-3G basis set with respect to the RHF energy against DiffiQult. In the column ``Parameters,'' $\alpha$, $d$, and $\bm{R}$ respectively represent all exponent coefficients, all contraction coefficients, and all center positions of the GTOs in the basis set. Each row of the sub-column ``Quantity'' shows the total number of parameters. The numbers in parentheses (when they appear) after them represent the reduced numbers of unique parameters after imposing parameter correlations in Quiqbox.}
    \label{tb:poc}
\end{table*}

Nevertheless, Quiqbox's optimization function showed a significant advantage against DiffiQult as presented in Table~\ref{tb:poc}. We included two testing cases for $\rm{H_2}$ (with the bond length at 0.7408 \AA): one was for optimizing only the exponent and contraction coefficients ($\alpha$ and $d$), and the other one also included the GTO centers ($\bm{R}$). The reason for doing so is that DiffiQult does not support optimizing GTO centers and the other two simultaneously, which is not a limitation for Quiqbox. Using the same numerical optimizer as DiffiQult, BFGS~\cite{head1985broyden} with a line search algorithm using the strong Wolfe (SW) conditions~\cite{doi:10.1137/1013035}, Quiqbox provided roughly a 700 times speed-up. By replacing the line search algorithm with the one used in the nonlinear conjugate gradient method proposed by Hager and Zhang (HZ)~\cite{hager2005new}, Quiqbox was able to achieve faster convergence in terms of the number of steps with a slightly lower wall-clock time speed up. We also include the test results using a limited-memory version of BFGS (L-BFGS)~\cite{liu1989limited} (memory size set to 20) combined with the HZ linear algorithm to demonstrate a more practical use case. When only using a first-order stochastic gradient descent optimizer like Adam~\cite{kingma2014adam}, Quiqbox still outperformed DiffiQult with over 100 times speed-up. 

The drastic improvement of basis set optimization's efficiency pushes forward the practicality of variational basis set optimization, especially for larger systems. For example, in the case of optimizing the STO-3G basis set for LiH (with the bond length at 1.5949 \AA), Quiqbox was able to finish the optimization within one hour. In contrast, DiffiQult did not achieve convergence even after running for more than 12 hours. This result is included in Table~\ref{tb:poc} as well.

\section{Supplementary information for atomization and formation energy computation}\label{app:reaction}
We herein provide the preliminary data used to compile Table~\ref{tb:reactions} in Sec.~\ref{sec:H2O2}. Specifically, the geometries of tested molecules are in Table~\ref{tb:geoinfo}, the size information of compared basis sets is in Table~\ref{tb:bsinfo}, and the ground-state HF energies (including nuclear repulsion) of the reactants and products used for the atomization and formation calculation are in Table~\ref{tb:groundstate}.

\begin{center}
    \begin{table}[htbp]
    \setlength{\tabcolsep}{5pt}
    \renewcommand{\arraystretch}{1.2}
    \centering
    \begin{tabular}{ccrrr}
    \toprule[1.5pt]
    \multicolumn{2}{c}{\multirow{2}{*}{Molecule}} & \multicolumn{3}{c}{Coordinate (\AA)} \\
    \multicolumn{2}{c}{} & \multicolumn{1}{c}{x} & \multicolumn{1}{c}{y} & \multicolumn{1}{c}{z} \\\midrule[1.2pt]\midrule[1.2pt]
    \multirow{2}{*}{$\rm H_2$} & H & 0.0000 & 0.0000 & 0.0000 \\
    & H & 0.7122 & 0.0000 & 0.0000 \\\cline{1-5}
    \multirow{2}{*}{$\rm O_2$} & O & 0.0000 & 0.0000 & 0.0000 \\
    & O & 1.2172 & 0.0000 & 0.0000 \\\cline{1-5}
    \multirow{3}{*}{$\rm H_2O$} & O & 0.0000 & 0.0000 & 0.1272 \\
    & H & 0.0000 & 0.7581 & -0.5086 \\
    & H & 0.0000 & -0.7581 & -0.5086 \\\cline{1-5}
    \multirow{4}{*}{$\rm H_2O_2$} & O & 0.0000 & 0.6981 & -0.0504 \\
    & O & 0.0000 & -0.6981 & -0.0504 \\
    & H & 0.8712 & 0.8912 & 0.4034 \\
    & H & -0.8712 & -0.8912 & 0.4034\\
     \bottomrule[1.5pt]
    \end{tabular}
    \caption{The theoretical equilibrium geometries of various molecules according to the HF approximation using STO-3G~\cite{cccbdb}.}
    \label{tb:geoinfo}
    \end{table}
\end{center}

\begin{center}
    \begin{table}[htbp]
    \setlength{\tabcolsep}{5pt}
    \renewcommand{\arraystretch}{1.2}
    \centering
        \begin{tabular}{lcccc}
            \toprule[1.5pt]
            \multirow{2}{*}{Basis   Set} & \multicolumn{2}{c}{H} & \multicolumn{2}{c}{O} \\ \cline{2-5} 
             & $W$ & $n_{\rm GTO}$ & $W$ & $n_{\rm GTO}$ \\
             \midrule[1.2pt]\midrule[1.2pt]
            STO-3G & \multirow{2}{*}{1} & \multirow{2}{*}{3} & \multirow{2}{*}{5} & \multirow{2}{*}{15} \\
            STO-3G-opt &  &  &  & \\\cline{1-5}
            STO-3G** & \multirow{2}{*}{4} & \multirow{2}{*}{6} & \multirow{2}{*}{11} & \multirow{2}{*}{21} \\
            STO-3G-opt** &  &  &  & \\\cline{1-5}
            6-31G & 2 & 4 & 9 & 22 \\
            cc-pVDZ & 5 & 8 & 15 & 40\\
            \bottomrule[1.5pt]
        \end{tabular}
        \caption{The size information of the basis sets tested in Sec.~\ref{sec:H2O2}. Specifically, the total number of basis functions ($W$) and primitive Cartesian GTOs ($n_{\rm GTO}$) used to build each basis set are included.}
    \label{tb:bsinfo}
    \end{table}
\end{center}

\begin{center}
    \begin{table*}[htbp]
    \resizebox{\textwidth}{!}{%
    \setlength{\tabcolsep}{4pt}
    \renewcommand{\arraystretch}{1.2}
    \centering
    \begin{tabular}{lclclclclclcl}
        \toprule[1.5pt]
        \multicolumn{1}{l}{\multirow{2}{*}{Basis   Set}} & \multicolumn{2}{c}{H/UHF (\hspace{-0.1em}$M_s\hspace{-0.25em}=\hspace{-0.25em}\frac{1}{2}$\hspace{-0.1em})} & \multicolumn{2}{c}{H$_2$/RHF} & \multicolumn{2}{c}{O/UHF (\hspace{-0.1em}$M_s\hspace{-0.25em}=\hspace{-0.25em}1$\hspace{-0.1em})} & \multicolumn{2}{c}{O$_2$/UHF (\hspace{-0.1em}$M_s\hspace{-0.25em}=\hspace{-0.25em}1$\hspace{-0.1em})} & \multicolumn{2}{c}{H$_2$O/RHF} & \multicolumn{2}{c}{H$_2$O$_2$/RHF} \\ \cline{2-13} 
        \multicolumn{1}{c}{} & \multicolumn{1}{c}{$E_{\rm tot}$ (Ha)} & \multicolumn{1}{r}{Error} & \multicolumn{1}{c}{$E_{\rm tot}$ (Ha)} & \multicolumn{1}{r}{Error} & \multicolumn{1}{c}{$E_{\rm tot}$ (Ha)} & \multicolumn{1}{r}{Error} & \multicolumn{1}{c}{$E_{\rm tot}$ (Ha)} & \multicolumn{1}{r}{Error} & \multicolumn{1}{c}{$E_{\rm tot}$ (Ha)} & \multicolumn{1}{r}{Error} & \multicolumn{1}{c}{$E_{\rm tot}$ (Ha)} & \multicolumn{1}{r}{Error} \\ \midrule[1.2pt]\midrule[1.2pt]
        STO-3G       & -0.4666 & 6.68\% & -1.1175 & 1.39\% & -73.8042 & 1.36\% & -147.6364 & 1.37\% & -74.9659 & 1.44\% & -148.7650 & 1.38\% \\
        STO-3G-opt   & -0.4970 & 0.60\% & -1.1262 & 0.62\% & -74.3185 & 0.67\% & -148.5990 & 0.73\% & -75.5205 & 0.71\% & -149.7905 & 0.70\% \\
        STO-3G**     & -0.4666 & 6.68\% & -1.1212 & 1.06\% & -73.8197 & 1.34\% & -147.7263 & 1.31\% & -75.0161 & 1.38\% & -148.8539 & 1.32\% \\
        STO-3G-opt** & -0.4970 & 0.60\% & -1.1264 & 0.60\% & -74.3599 & 0.61\% & -148.7523 & 0.63\% & -75.5831 & 0.63\% & -149.9181 & 0.62\% \\
        6-31G        & -0.4982 & 0.36\% & -1.1266 & 0.58\% & -74.7803 & 0.05\% & -149.5450 & 0.10\% & -75.9797 & 0.11\% & -150.7024 & 0.10\% \\
        6-31G-opt    & -0.4993 & 0.14\% & -1.1307 & 0.22\% & -74.7858 & 0.04\% & -149.5619 & 0.09\% & -75.9975 & 0.09\% & -150.7245 & 0.08\% \\
        cc-pVDZ      & -0.4993 & 0.14\% & -1.1278 & 0.48\% & -74.7923 & 0.04\% & -149.6278 & 0.04\% & -76.0235 & 0.05\% & -150.7824 & 0.04\% \\
        CBS Limit    & -0.5000 & 0.00\% & -1.1332 & 0.00\% & -74.8194 & 0.00\% & -149.6900 & 0.00\% & -76.0627 & 0.00\% & -150.8489 & 0.00\% \\
        \bottomrule[1.5pt]
    \end{tabular}}
    \caption{The ground-state HF energies (including nuclear repulsion) of H, O, and related molecules (in a fixed molecular geometry shown in Table~\ref{tb:geoinfo}) using various basis sets. For each tested basis set, both the energy value and its relative error with respect to the extrapolated CBS limit~\cite{HALKIER1999437} (based on the results using cc-pVDZ, TZ, and QZ) are provided. UHF is applied to compute the open-shell ground states, for which the $z$ component of total spin ($M_s$) is also included in the table.}
    \label{tb:groundstate}
    \end{table*}    
\end{center}

\newpage

\section{Acronyms and notations}\label{app:notation}

\begin{table}[htp]
    \centering
    \setlength{\tabcolsep}{4.5pt}
    \renewcommand{\arraystretch}{1.2}
    \begin{tabular}{rllll}
    \toprule[1.5pt]
    Acronym & Full name                 \\ \midrule[1.2pt]\midrule[1.2pt]
    AO      & Atomic orbital.            \\
    GTO     & Gaussian-type orbital.     \\
    CGTO    & Contracted GTO.            \\
    MCGTO   & Mixed-contracted GTO.      \\\bottomrule[1.5pt]
    \end{tabular}
    \caption{The acronyms of single-particle spatial wavefunctions.}
    \label{tb:Oname}
\end{table}    
\begin{table}[htp]
    \centering
    \setlength{\tabcolsep}{4.5pt}
    \renewcommand{\arraystretch}{1.2}
    \begin{tabular}{rp{0.36\textwidth}}
    \toprule[1.5pt]
    Symbol           & Meaning                                                     \\ \midrule[1.2pt]\midrule[1.2pt]
    $N$              & Number of electrons.                                         \\
    $N^{\uparrow}$   & Number of spin-up electrons.                                 \\
    $N^{\downarrow}$ & Number of spin-down electrons.                               \\
    $N'$             & Number of nuclei.                                            \\
    $\mathbb{M}$     & Set of nuclear positions and the corresponding charges.      \\
    $H_{elec}$       & Electronic Hamiltonian.                                      \\
    $H_{nuc}$        & Sum of nuclear kinetic and nuclear repulsion energy. \\
    $H_{tot}$        & Total Hamiltonian.                                           \\
    $\mathcal{E}_0$    & True ground-state energy of $H_{elec}$.                      \\\bottomrule[1.5pt]
    \end{tabular}
    \caption{The notations of the physical variables determined by a many-electron system.}
    \label{tb:MEsystem}
\end{table}
\begin{table}[htp]
    \centering
    \setlength{\tabcolsep}{4.5pt}
    \renewcommand{\arraystretch}{1.2}
    \begin{tabular}{rp{0.3\textwidth}}
    % \begin{tabular}{rp{0.75\textwidth}}
    \toprule[1.5pt]
    Symbol                         & Meaning                                                                    \\ \midrule[1.2pt]\midrule[1.2pt]
    $\phi_n$                       & The $n$th basis function.                                      \\
    $\bm{S}$                       & The overlap matrix of basis set $\{\phi_n \,|\,  n\!=\!1,\dots,W\}$.                \\
    % $W$                            & Total number of basis functions                                            \\
    $\varphi_m$                    & The $m$th orthonormalized basis function.                                            \\
    $\Psi_0$                       & Trial wavefunction of the approximate ground state based on an ansatz.             \\
    $\Phi_n$                       & The $n$th eigenfunction of the electronic Hamiltonian.                               \\
    $\bm{c}$                       & Ansatz parameters of $\Psi_0$.                                              \\
    $\tilde{E_0}$                  & Energy expectation with respect to $\Psi_0$.                            \\
    $E_0$                          & Lower bound of $\tilde{E_0}$.                                               \\
    $\bm{A}$                       & One-electron integral tensor of $\{\varphi_m\}$.               \\
    $\bm{B}$                       & Two-electron integral tensor of $\{\varphi_m\}$.               \\
    $\bm{c^o}$                     & Value of  $\bm{c}$ that minimizes $\tilde{E_0}$ to $E_0$.                   \\
    $\alpha$                       & The exponent coefficient of a GTO.                                                      \\
    $\mathpzc{l}\!\equiv\!\mathpzc{i}\!+\!\mathpzc{j}\!+\!\mathpzc{k}$ & The orbital angular momentum of a GTO.\\
    $\bm{R}\!\equiv\![X,\, Y,\, Z]$& The center position of a GTO.                                        \\
    $d$                            & The contraction coefficient in a CGTO.                                          \\
    $\bm{\theta}\:\; (\{\theta_i\})$ & Basis set parameters.                                                       \\
    $\{p_j^n\}$                    & All the differentiable CGTO parameters of the $n$th basis function.                        \\
    $\{p_j\}$                      & All the differentiable CGTO parameters in the basis set.                                   \\
    $\psi_n^{\theta_i}$            & The basis function equivalent to the partial derivative of $\phi_n$ with respect to $\theta_i$.\\
    $\mathbb{D} $ & The derivative basis set that transforms the parameter gradients of elements in $\{\phi_n\}$ into electronic integrals.\\\bottomrule[1.5pt]
    
    \end{tabular}
    \caption{The notations of the variables and quantities related to basis functions.}
    \label{tb:Basis}
\end{table}

\end{appendices}

\end{document}